%% file: ms.tex
\documentclass[iop]{emulateapj}

\usepackage[colorlinks=true,linkcolor=blue,anchorcolor=blue,citecolor=blue,urlcolor=blue]{hyperref}  
\hypersetup{pdfauthor={Ho Seong HWANG}, pdftitle={SHELS: WISE 22 $\mu$m-selected galaxies}}

\newcommand{\kmsb}{km s$^{-1}$}

\newcommand{\irasb}{\textit{IRAS}}
\newcommand{\iras}{\textit{IRAS} }
\newcommand{\akari}{\textit{AKARI} }
\newcommand{\wise}{\textit{WISE} }

\newcommand{\spitzer}{\textit{Spitzer} }
\newcommand{\herschel}{\textit{Herschel Space Observatory} }

\slugcomment{Last updated: \today}
\shorttitle{SHELS: \wise 22 $\mu$m-selected galaxies}
\shortauthors{Hwang et al.}

\usepackage{natbib}

\begin{document}

\title{SHELS: Optical Spectral Properties of WISE 22 $\mu$m-selected Galaxies}

\author{Ho Seong Hwang\altaffilmark{1}}
\author{Margaret J. Geller\altaffilmark{1}}
\author{Michael J. Kurtz\altaffilmark{1}} 
\author{Ian P. Dell'Antonio\altaffilmark{2}} 
\author{Daniel G. Fabricant\altaffilmark{1}} 

\altaffiltext{1}{Smithsonian Astrophysical Observatory, 60 Garden St., Cambridge, MA 02138;
hhwang@cfa.harvard.edu, mgeller@cfa.harvard.edu, mkurtz@cfa.harvard.edu, dfabricant@cfa.harvard.edu}
\altaffiltext{2}{Department of Physics, Brown University, Box 1843, Providence, RI 02912;
ian@het.brown.edu}

\begin{abstract}
We use a dense, complete redshift survey,
   the Smithsonian Hectospec Lensing Survey (SHELS), 
   covering a 4 square degree region of
   a deep imaging survey, the Deep Lens Survey (DLS),
   to study the optical spectral properties of
   Wide-field Infrared Survey Explorer ({\it WISE}) 22 $\mu$m-selected galaxies.
Among 507 \wise 22 $\mu$m-selected sources 
  with (S/N)$_{22\mu{\rm m}}\geq3$ ($\approx S_{22\mu m}\gtrsim2.5$ mJy),
  we identify the optical counterparts of 481 sources ($\sim98\%$) at 
  $R<25.2$ in the very deep, DLS $R$-band source catalog.
Among them, 
  337 galaxies at $R<21$ have SHELS spectroscopic data.
Most of these objects are at $z<0.8$.
The infrared (IR) luminosities are in the range
 $4.5\times10^8  ({\rm L}_\odot) \lesssim L_{IR} \lesssim 5.4\times10^{12}  ({\rm L}_\odot)$.
Most 22 $\mu$m-selected galaxies are
  dusty star-forming galaxies with a small ($<$1.5) 4000 \AA~break.
The stacked spectra of the 22 $\mu$m-selected galaxies binned in IR luminosity
   show that the strength of the [O III] line relative to H$\beta$
   grows with increasing IR luminosity.
The optical spectra of the 22 $\mu$m-selected galaxies
  also show that there are some ($\sim2.8\%$) unusual galaxies with 
  very strong [Ne III] $\lambda$3869, 3968 emission lines
  that require hard ionizing radiation such as AGN or extremely young massive stars.
The specific star formation rates (sSFRs) 
  derived from the 3.6 and 22 $\mu$m flux densities
  are enhanced if the 22 $\mu$m-selected galaxies 
  have close late-type neighbors.
The sSFR distribution of the 22 $\mu$m-selected galaxies containing
  active galactic nuclei (AGNs)
  is similar to the distribution for 
  star-forming galaxies without AGNs.
We identify 48 dust-obscured galaxy (DOG) candidates 
  with large ($\gtrsim1000$) mid-IR to optical flux density ratio.
The combination of deep photometric and spectroscopic data with \wise data
  suggests that \wise can probe the universe to $z\sim2$.
\end{abstract}


\section{Introduction}

The {\it Infrared Astronomical Satellite} (\iras; \citealt{neu84})
  and the \akari satellite \citep{mur07} conducted 
  infrared (IR) all-sky surveys; they 
  detected a tremendous number of IR sources over the entire sky.
Recent IR satellites including \herschel \citep{pil10}
  have also carried out several small-area survey programs;
  they extended our understanding of the IR universe to high redshift
  (see \citealt{gc00, soi08, elb11, lutz11, oli12} for a review).
  
Wide-field galaxy redshift surveys
  \citep{huc83,del86,york00,col01,jones04,dri11} enlarge our view
  of the universe by adding the third dimension (i.e. redshift) to the projected sky
  (e.g., \citealt{gh89, gott05}).

The combination of ``wide-field'' IR and redshift surveys thus
  provides valuable, large samples of IR-detected galaxies
  without ambiguity in projection along the line of sight
  (e.g., \citealt{cao06,hwa07,hou09,wr09cat}).
These samples are an important basis for studying
  the optical/IR properties of IR bright galaxies
  and their environmental dependence
  (e.g., \citealt{goto05ir,gel06,hwa10lirg,hwa10tdust,jhlee10mul,leejc11ulirg,goto11lfsdss,dar11,don12}).
There are also several redshift surveys {\it dedicated} to IR sources
  (see \citealt{sm96} and \citealt{soi08} and references therein;
   see also \citealt{lake12}).

\input{table1}

The IR bright galaxies are mainly powered by
  star formation (often they are also powered by  active galactic nuclei (AGNs)).
To understand the physical processes responsible for the IR activity,
  it is thus important to understand the physical parameters
  affecting the star formation rates (SFRs) of galaxies
  (see \citealt{mo07,ken12} for a review).
The galaxy stellar mass is one of the important parameters.
Empirically, the SFR per unit stellar mass (i.e. specific SFR, sSFR) 
  is a useful indicator of SF activity (SFA) of galaxies 
  \citep{bri04,dad07,pan09,gmag10sfrm,elb11}.

\begin{figure*}
\center
\includegraphics[width=140mm]{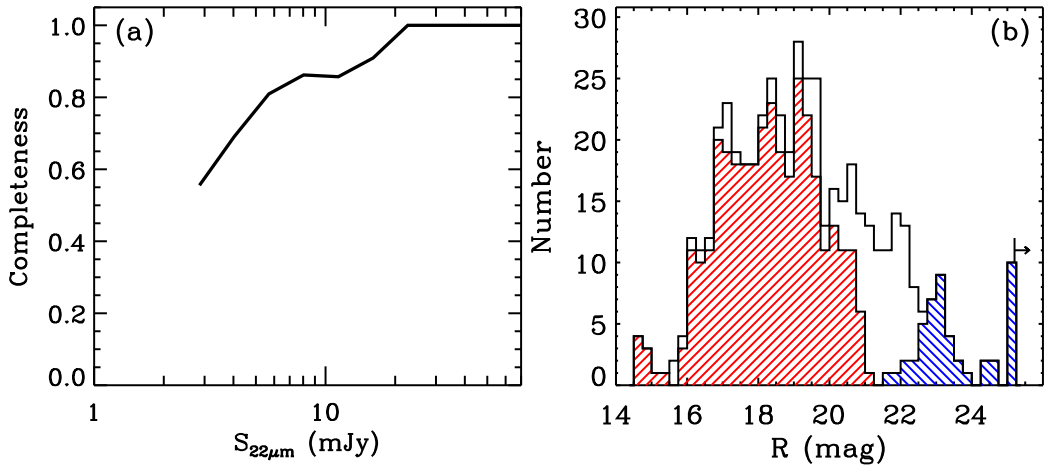}
\caption{Spectroscopic completeness
  for \wise 22 $\mu$m sources
  as a function of 22 $\mu$m flux density (a).
$R$-band magnitude distribution for \wise 22 $\mu$m sources (b).
All 22 $\mu$m sources and
 the sources with spectroscopic redshifts
 are denoted by open and red hatched histograms, respectively.
The dust-obscured galaxy (DOG) candidates (see Section \ref{dogs})
  are indicated by a hatched histogram with
  orientation of 315$^\circ$ ($\setminus\setminus$ with blue color) 
  relative to horizontal.
The right most bin with a right arrow indicates the limiting
  $R$-band magnitude.
}\label{fig-comp}
\end{figure*}

The proximity and gas content of the nearest neighbor galaxy
  are also important parameters.
In the hierarchical picture of galaxy formation,
  it is evident that galaxy interactions and mergers
  strongly affect the SFA of galaxies over several billion years.
Observational studies
  clearly show that the sSFRs of IR bright galaxies are enhanced
  if the galaxies have gas-rich close neighbors
  both in low-$z$ and in high-$z$ universe 
  (e.g., \citealt{gel06,kar10,kar12,hwa10lirg,hwa11inter}).
The presence of AGN
  is also an important parameter affecting the SFRs of galaxies.
Recently, the interplay between AGN and SF within a galaxy
  is a matter of much debate: 
  is the relationship causal or non causal? (e.g., \citealt{ah12,mul12,fab12}).

Here, we discuss the optical spectral properties of IR-selected galaxies
  to improve understanding of the drivers of the SFA
  in star-forming galaxies at intermediate redshift.
We use a dense, complete redshift survey,
  the Smithsonian Hectospec Lensing Survey (SHELS; \citealt{gel05,gel10,gel12}),
  covering a 4 square degree region of
  a deep imaging survey, the Deep Lens Survey (DLS; \citealt{wit02, wit06}).
This region is uniformly covered by 
  the Wide-field Infrared Survey Explorer ({\it WISE}; \citealt{wri10})
  with excellent sensitivity 
  in the mid-IR (MIR) bands (3.4, 4.6, 12, and 22 $\mu$m).
\wise is 100 times more sensitive than \iras at 12 $\mu$m,
  and has a resolution (6\arcsec.5 at 12 $\mu$m) 
  much better than the arcmin resolution of \irasb.
  
Because SHELS is a magnitude-limited survey down to $R=20.6$ (mag)
  without any complex selection effects,
  the combination of SHELS and \wise  data
  provides a large, unbiased sample of {\it WISE}-selected galaxies
  with spectroscopic parameters including redshifts,
  4000 \AA~break and emission line fluxes.
       
In this study, we focus on the 22 $\mu$m-selected sample of SHELS galaxies
  with signal-to-noise ratio (S/N)$_{22\mu{\rm m}}\geq3$ 
 ($\approx S_{22\mu m}>2.5$ mJy).
We discuss 12 $\mu$m-selected galaxies
  in a companion paper (H. S. Hwang et al. 2012, in preparation).  
Section \ref{data} describes the observations.
We examine the optical and MIR properties of
   22 $\mu$m-selected SHELS galaxies in \S \ref{results},
   and conclude in \S \ref{sum}.
Throughout,
  we adopt flat $\Lambda$CDM cosmological parameters:
  $H_0 = 70$ km s$^{-1}$ Mpc$^{-1}$, $\Omega_{\Lambda}=0.7$ and $\Omega_{m}=0.3$.

\input{table2}

\input{table3}

\section{The Data}\label{data}

\subsection{SHELS}

SHELS is a magnitude-limited redshift survey
  covering the 4 square degree region of 
  the DLS F2 field \citep{gel05,gel10,gel12}.
The DLS is an NOAO key program covering 20 square degrees 
  in five separate $2\arcdeg\times2\arcdeg$ fields \citep{wit02, wit06}.
The F2 field is one of the five fields and is centered on 
  $\alpha = 09^{\rm h}19^{\rm m}32^{\rm s}.4$ and
  $\delta = + 30^\circ 00\arcmin 00\arcsec$.
The DLS $R$-band images were taken in seeing $0\arcsec.9$;
  the source detection on these images 
  reaches down to 28.7 mag per square arcsecond.
  
The spectroscopic observations for SHELS galaxies
  were conducted for galaxies down to $R=20.6$
  with the Hectospec 300 fiber spectrograph \citep{fab05} on the MMT
  (see \citealt{gel12} for detailed description of the spectroscopic observations).
We used Hectospec's 270 line mm$^{-1}$ grating 
  that provides a dispersion of 1.2 \AA~pixel$^{-1}$ and a resolution of $\sim$6 \AA.
The spectra cover the wavelength range $3650-9150$ \AA. 
Exposure times ranged from 0.75 to 2 hr.
We also observed some galaxies fainter than $R=20.6$
  when fibers were available.
The total number of objects with SHELS spectra is about 19,800
  including 2874 sources with $R=20.6-21$. 

The spectroscopic completeness is 98\% at $R\leq20.3$,
  and the differential completeness for $R=20.3-20.6$ is 89\%.
The objects missed in the spectroscopic observations at $R<20.6$
  are those near the survey edges, those near bright stars, or point sources.

\subsection{{\it WISE}}\label{wise}

We use the \wise all-sky survey catalog\footnote{http://wise2.ipac.caltech.edu/docs/release/allsky/expsup/},
  containing uniform photometric data for over 563 million objects at 4 MIR bands 
  (3.4, 4.6, 12 and 22 $\mu$m).
\wise covers the entire region of the SHELS field to a homogeneous depth.
We use the point source profile-fitting magnitudes,
  and restrict our analysis to the sources with S/N$\geq3$ at 22 $\mu$m.
The \wise 5$\sigma$ photometric sensitivity is estimated to be better 
  than 0.08, 0.11, 1 and 6 mJy
  at 3.4, 4.6, 12 and 22 $\mu$m
  in unconfused regions on the ecliptic plane \citep{wri10}.

After rejecting bright stars
  through visual inspection of their optical images,
  we have 507 sources with 22 $\mu$m detection ($S/N\geq3$)
  in the SHELS field.
We cross-correlate these IR sources with the objects 
  in the DLS $R$-band source catalog (down to $R<25.2$)
  with a matching tolerance of 3\arcsec($\sim$ 0.5$\times$FWHM of the \wise PSF at 3.4 $\mu$m).
We identify 481 matches and 26 sources without optical counterparts.
Among the 481 matches,
  there are 337 (107) galaxies with $S/N_{22{\mu m}}\geq3$ ($\geq5$) 
  and a measured redshift in SHELS.
We discuss the sources without optical counterparts in Section \ref{dogs}.

Table \ref{tab-stat} summarizes the statistics for the number
  of \wise 22 $\mu$m sources in our sample.
Figure \ref{fig-comp} shows the spectroscopic completeness
  for \wise 22 $\mu$m sources (left panel).
The completeness is 100\%  at $\gtrsim20$ mJy, and
  reaches $\sim60\%$ even at the faintest level of $\sim3$ mJy.
We also show the $R$-band magnitude distribution 
  of the 22 $\mu$m sources in the right panel.
As expected, most 22 $\mu$m sources at $R<20.6$
  have spectroscopic redshifts (red hatched histogram).

We list 337 galaxies with $S/N_{22{\mu m}}\geq3$ 
  and spectroscopic redshifts in Table \ref{tab-samp}
  with their optical properties including
  $R$-band magnitude, redshift, $D_n4000$, and SF/AGN classification.
We also list the \wise properties of these galaxies
  including the IR luminosity derived from the 22 $\mu$m flux density
  and the stellar mass from the 3.4 $\mu$m flux density (see Section \ref{mass})
  in Table \ref{tab-wise}.

\begin{figure}
\center
\includegraphics[width=80mm]{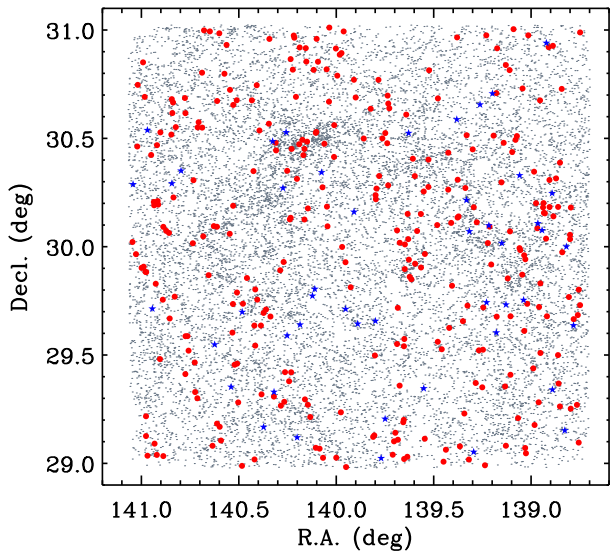}
\caption{Spatial distribution of
  SHELS galaxies on the sky.
The red circles are \wise 22 $\mu$m-selected galaxies,
  and gray dots are all SHELS galaxies regardless of \wise detection.
Blue star symbols are DOG candidates (see Section \ref{dogs}).
}\label{fig-spat}
\end{figure}
  
We show the spatial distribution of the 22 $\mu$m-selected SHELS galaxies (red circles)
  in company with all SHELS galaxies regardless of \wise detection (gray dots)
  in Figure \ref{fig-spat}.
Some galaxy clusters are obvious in the distribution of all SHELS galaxies
  on the sky (see \citealt{gel10} for a list of candidate clusters),
  but the 22 $\mu$m-selected galaxies are less strongly clustered.
These 22 $\mu$m-selected galaxies are mainly star-forming galaxies;
  thus they tend to avoid the core region of clusters as expected
  (e.g., \citealt{ph09}).

\begin{figure}
\center
\includegraphics[width=80mm]{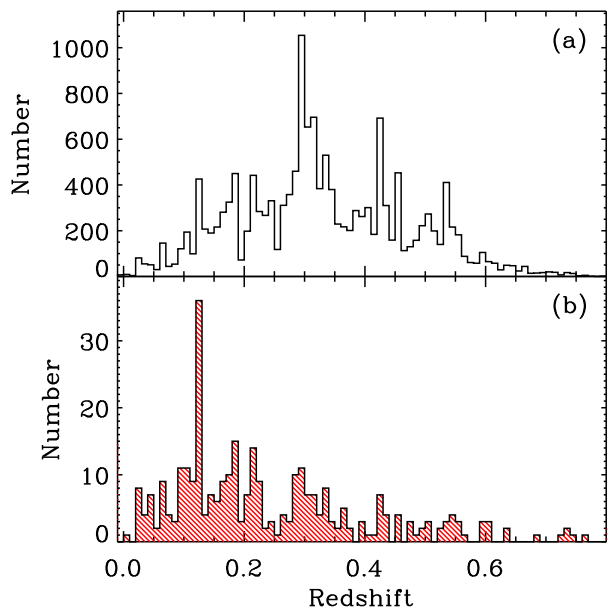}
\caption{Redshift distribution of all SHELS galaxies (a)
  and of 22 $\mu$m-selected galaxies (b).
}\label{fig-zhist}
\end{figure}
  
Figure \ref{fig-zhist} shows
  the redshift distribution of all SHELS galaxies (a) and
  of \wise 22 $\mu$m-selected galaxies (b).
The redshift distribution of all SHELS galaxies in panel (a) shows
  dominant peaks around $z\sim0.3$ and 0.4, 
  indicating the presence of galaxy clusters \citep{gel10}.
Interestingly, there is a prominent peak
  for 22 $\mu$m-selected galaxies in panel (b)
  at $z\sim0.13$.
This peak corresponds to a galaxy group (see \citealt{gel10}). 
However, the 22 $\mu$m-selected galaxies are not centrally concentrated,
  but are in the outskirts of the group.

\begin{figure*}
\center
\includegraphics[width=140mm]{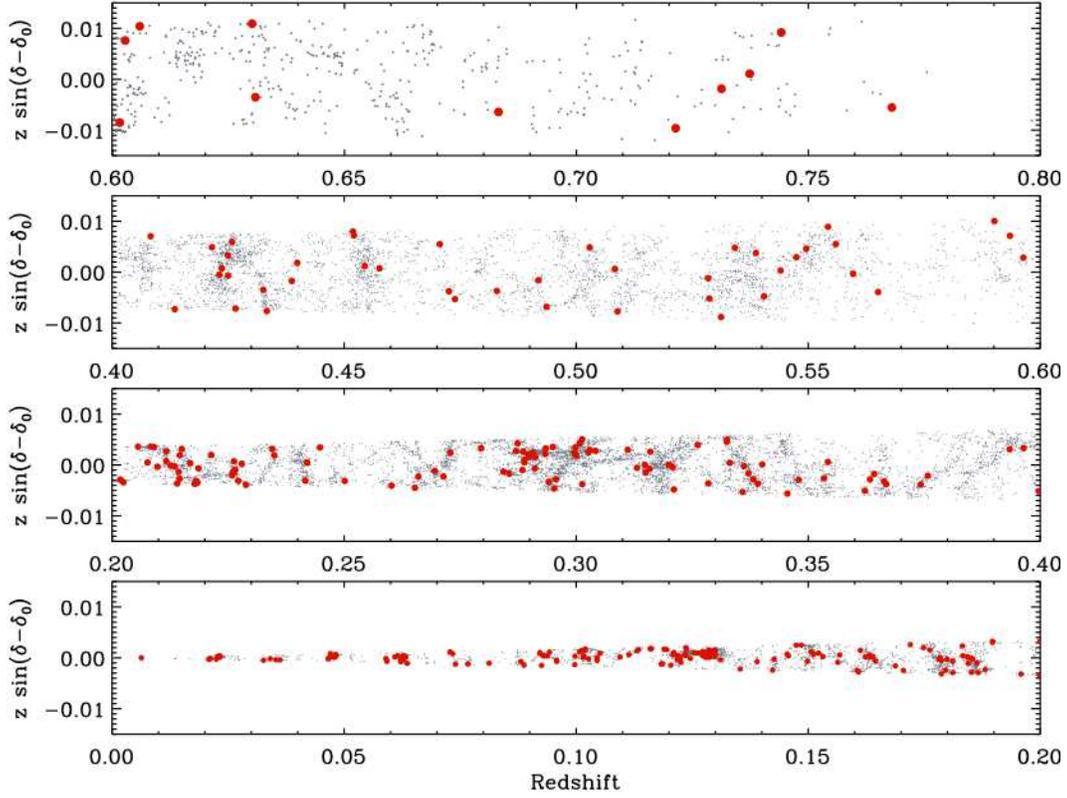}
\caption{Redshift cone diagram for
  SHELS galaxies projected along the declination direction.
We project the entire right ascension range onto 
  the declination-redshift plane.
The red circles are \wise 22 $\mu$m-selected galaxies,
  and gray dots are all SHELS galaxies regardless of \wise detection.
}\label{fig-wedge}
\end{figure*}

Figure \ref{fig-wedge} shows a cone diagram
  projected along the declination direction.
We project the entire right ascension range onto 
  the declination-redshift plane.
The red circles are 22 $\mu$m-selected galaxies,
  and the gray dots are all SHELS galaxies regardless of \wise detection.
There are detections up to $z\sim0.77$
  (there are also 20 galaxies/quasars at $z=0.8-3.0$ not shown in the figure). 
The figure shows again that the 22 $\mu$m-selected galaxies
  avoid the highest density regions as expected.

\begin{figure}
\center
\includegraphics[width=80mm]{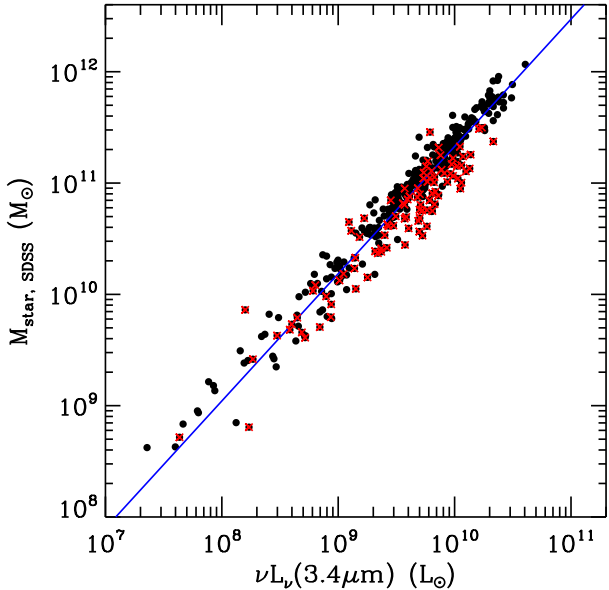}
\caption{Stellar mass from the SDSS database vs.
  rest-frame 3.4 $\mu$m luminosity for SHELS galaxies regardless of 22 $\mu$m detection (black circles).
The AGN contribution to the 3.4 $\mu$m luminosity is removed with the SED fit.
The solid line is the best fit with an ordinary least-squares bisector method 
  (see equation (\ref{eq-mass})).
Red crosses are 22 $\mu$m-selected galaxies.
}\label{fig-mass}
\end{figure}

\subsection{Stellar Masses and SFRs of \wise 22 $\mu$m-selected Galaxies}\label{mass}

\wise 3.4 $\mu$m data probe the old stellar components
  dominating the stellar mass of galaxies.
Thus the rest-frame luminosity around this wavelength 
  can be a useful tracer of stellar mass (e.g., \citealt{han07, ko12, lin12}).
There are other components contributing at this wavelength
  including 3.3 $\mu$m polycyclic aromatic hydrocarbon (PAH) emission
  and AGN dust emission \citep{lu03,fla06,bmag08,leejc12akari},
  but they seem to have little effect on the correlation
  between stellar mass and 3.4 $\mu$m rest-frame luminosity (see also \citealt{dra07}).
To determine the correlation between stellar mass and 3.4 $\mu$m rest-frame luminosity,
  we plot stellar masses derived from the SDSS data
  against 3.4 $\mu$m rest-frame luminosities
  for SHELS galaxies 
  in Figure \ref{fig-mass}.
We use stellar mass estimates 
  for the spectroscopic sample of SDSS galaxies
  from the MPA/JHU DR7 value-added galaxy 
  catalog\footnote{http://www.mpa-garching.mpg.de/SDSS/DR7/Data/stellarmass.html}
  (VAGC).
These estimates are based on the fit of SDSS five-band photometry 
  to the models of \citet{bc03} (see also \citealt{kau03}).
We convert the stellar masses in the MPA/JHU DR7 VAGC
  that are based on the Kroupa IMF \citep{kro01}
  to those with Salpeter IMF \citep{sal55} by dividing them by a factor of 0.7 \citep{elb07}.

To compute the rest-frame luminosity in the \wise bands,
  we use the fitting code and the templates in \citet{ass10}.
The code utilizes a set of low-resolution empirical spectral energy distribution (SED) templates
  covering the wavelengths $0.03-30$ $\mu$m.
The templates include
  an old stellar population,
  a continuously star-forming population,
  a starburst population, and AGNs with varying amounts of
  reddening and absorption by the intergalactic medium.
We apply this code
  to our combined data set of SDSS $ugriz$, 
  KPNO $R$, \wise 3.4, 4.6, 12 and 22 $\mu$m photometry.
To remove the AGN contribution from the 3.4 $\mu$m luminosity,
  we estimate the AGN contribution to the 3.4 $\mu$m luminosity 
  based on the \citeauthor{ass10} templates.
The median AGN contribution to the 3.4 $\mu$m luminosity
  for the 22 $\mu$m-selected galaxies is only 6\%.
For each galaxy, we subtract the AGN contribution from the 3.4 $\mu$m luminosity 
  and use this quantity to derive a correlation with SDSS mass estimate.

\begin{figure}
\center
\includegraphics[width=85mm]{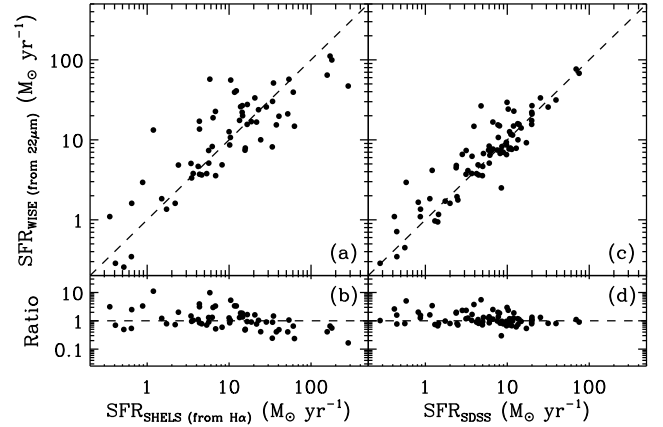}
\caption{Comparison of SFRs from \wise 22$\mu$m flux density (SFR$_{\rm WISE}$)
  with those from SHELS MMT spectra (SFR$_{\rm SHELS}$, a-b) 
  and from SDSS spectra (SFR$_{\rm SDSS}$, c-d).
We plot only galaxies with S/N$_{H\alpha}\ge20$.
The dashed line in each panel indicates the one-to-one relation.
}\label{fig-sfr}
\end{figure}

\begin{figure*}
\center
\includegraphics[width=170mm]{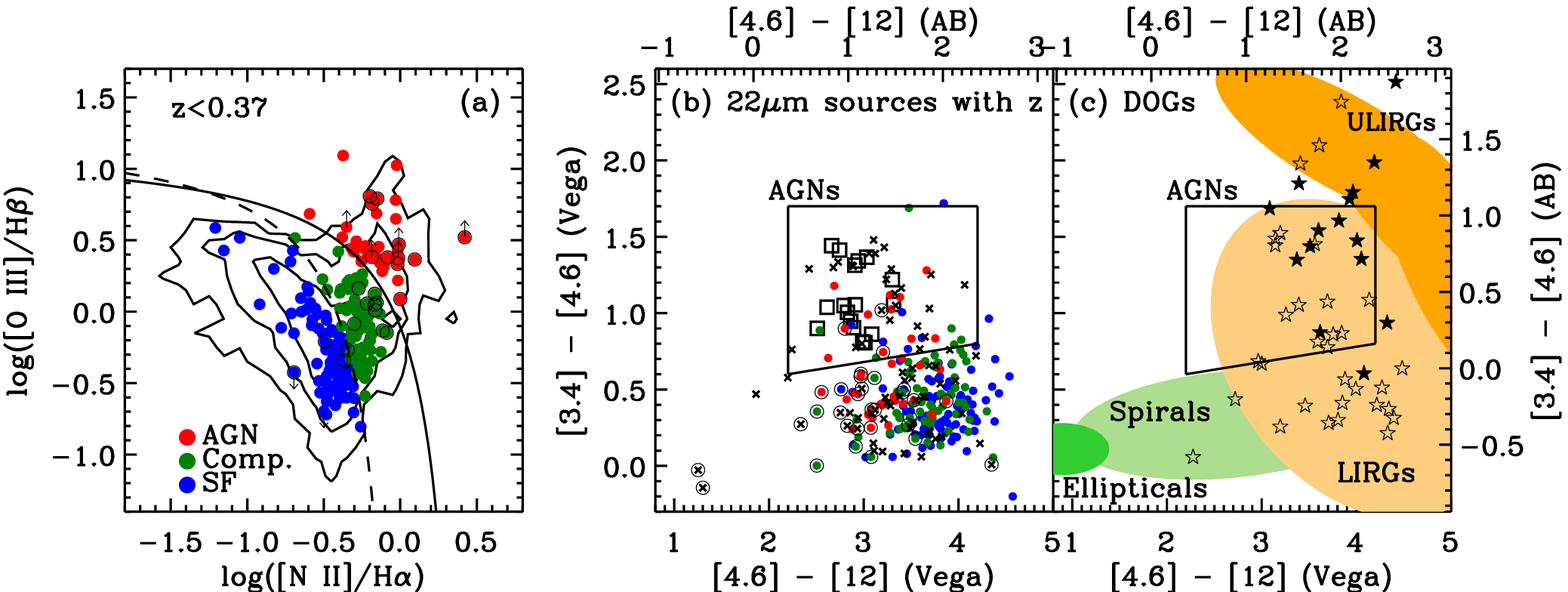}
\caption{AGN diagnostic diagrams for galaxies
   based on optical [O III]/H$\beta$ vs. [N II]/H$\alpha$ line ratios ({\it a})
   and on {\it WISE} colors ({\it b-c}). 
The solid and dashed lines in (a) indicate the extreme starbursts \citep{kew01}
  and pure SF limits \citep{kau03agn}, respectively.
The contours indicate the distribution of all SHELS galaxies 
  regardless of \wise detection.
The lowest contour level encloses 95\% of SHELS galaxies.
Different colored symbols represent different classes
  (AGN: red, Composite: green, SF: blue).
Galaxies with $S/N<3$ for H$\beta$ and
  for [O III] are shown as up and down arrows, respectively.
Galaxies with $D_n4000\geq1.5$ are represented by black open circles.
We plot only galaxies at $z<0.37$.
In (b), we plot all 22 $\mu$m-selected SHELS galaxies with spectroscopic redshifts.
Crosses in (b) are galaxies not classified in panel (a).
Open squares are broad-line AGNs, and
  open circles are galaxies with $D_n4000\geq1.5$.
Panel (c) shows {\it WISE} colors for DOG candidates 
  (filled star: $S/N\geq3$ at all bands, open star: $S/N<3$ at least one band).
Solid lines in (b-c) are the MIR AGN selection criteria proposed by 
  \citet{jar11}, respectively.
We mark several regions occupied by interesting classes of objects
 (ellipticals, spirals, LIRGs and ULIRGs; see Figure 12 in \citealt{wri10}).
 }\label{fig-agn}
\end{figure*}
  
Figure \ref{fig-mass} demonstrates the tight correlation
  between 3.4 $\mu$m rest-frame luminosities and 
  stellar masses derived from the SDSS data,
  suggesting that 3.4 $\mu$m rest-frame luminosity is indeed
  a good tracer of stellar mass (once AGN contribution is removed).
We use the bisector method \citep{iso90} to fit the correlation, and
  obtain the relation,
\begin{equation}
{\rm log}(M_{\rm star}/M_\odot) = {\rm log}(\nu L_\nu (3.4 \mu m)/L_\odot) \times (1.14\pm0.02) - (0.11 \pm0.17).
\label{eq-mass}
\end{equation}
\noindent We use this equation to compute stellar masses 
  for all of the SHELS 22 $\mu$m-selected galaxies
  except the broad-line AGNs where the AGN component dominates the SED.

We compute the IR luminosities ($L_{\rm IR}$) for 22 $\mu$m-selected galaxies
  from their 22 $\mu$m flux densities
  because 22 $\mu$m data
  are closer to the peak of IR emission than the other \wise bands and 
  because they are less affected by PAH emission features.
We use the SED templates of \citet[]{ce01} 
  as a basis for the SED fit.
IR luminosities extrapolated from a single passband
  have been examined in many papers;
  they agree very well with those measured with all FIR bands
  even in the presence of AGNs \citep{elb10, elb11}.
They also agree with the optical spectra \citep{hwa12a2199}.
However, it should be noted that 
  the SFRs  converted from IR luminosities
  for AGN-host galaxies (see Section \ref{starb})
  indicate a upper limit on the SFRs of their host galaxies.
An SED analysis with data points at $\lambda>22\mu$m is necessary 
  to estimate the AGN contribution at 22 $\mu$m correctly.

As a consistency check, 
  we compare SFRs converted from IR luminosities (SFR$_{\rm WISE}$) with 
   those from MMT and SDSS optical spectra (SFR$_{\rm SHELS}$ and SFR$_{\rm SDSS}$)
  in Figure \ref{fig-sfr}.
We convert the IR luminosity into SFR$_{\rm WISE}$ 
  using the relation in \citet{ken98} with the assumption of 
  a Salpeter IMF:
  SFR$_{\rm WISE}$ ($M_\odot$ yr$^{-1}$) $= 1.72\times10^{-10}L_{\rm IR} (L_\odot)$. 
We adopt SFR$_{\rm SHELS}$ from \citet{wes10},
  who compute the aperture and extinction-corrected SFRs
  from the H$\alpha$ fluxes derived from the MMT spectra.
A typical aperture correction factor defined by $A = 10^{-0.4 (m_{\rm total} - m_{\rm fiber})}$
  for 22 $\mu$m-selected galaxies is 4.8 where 
  $m_{\rm total}$ and $m_{\rm fiber}$ are total and fiber $R$-band magnitudes, respectively
  (see Fig. 2 in \citealt{wes10}).
A typical $V$-band extinction $A_V$ with \citet{cal00} reddening law
  for 22 $\mu$m-selected galaxies is 2.0 (see Fig. 3 in \citealt{wes10}), 
  consistent with the values for similar MIR-selected galaxies (e.g., see Fig. 11 in \citealt{cap08}).
The SFR$_{\rm SDSS}$ is from the MPA/JHU DR7 VAGC \citep{bri04},
  which provides the extinction and aperture corrected SFRs. 
We convert the SFR$_{\rm SDSS}$ in MPA/JHU DR7 VAGC
  based on the Kroupa IMF
  to the Salpeter IMF by dividing the SFR$_{\rm SDSS}$ by a factor of 0.7.

The comparison of SFR$_{\rm WISE}$ with 
  SFR$_{\rm SHELS}$ and SFR$_{\rm SDSS}$
  shows that SFR$_{\rm WISE}$ agrees well overall
  with the measurements from the optical spectra,
  demonstrating the consistency between the two measurements.
Scatter in the correlation results from uncertain 
  aperture and reddening corrections.
The SHELS galaxies in the left panel ($\langle z \rangle\sim0.14$)
  are on average more distant than
  the SDSS galaxies in the right panel ($\langle z \rangle\sim0.1$).
The fiber diameter for SHELS spectra (1\arcsec.5) is
  smaller than that for SDSS spectra (3\arcsec).
The scatter is thus slightly larger for SHELS galaxies.

\subsection{AGN Selection}\label{agnsel}

To identify AGN host galaxies among \wise 22 $\mu$m-selected galaxies,
  we use the Baldwin-Phillips-Terlevich (BPT) line ratio diagram based on 
  [O III]/H$\beta$ and [N II]/H$\alpha$ 
  (\citealt{bpt81,vo87}).
We apply this method only to galaxies at $z<0.37$ 
  because the H$\alpha$ line for galaxies at $z>0.37$
  is redward of the Hectospec spectral coverage ($3650-9150$ \AA).
We adopt line flux measurements from \citet{wes10},
  who provide the emission-line fluxes 
  from the continuum subtracted spectra of SHELS galaxies
  based on stellar population synthesis models of \citet{bc03}.
  
\begin{figure*}
\center
\includegraphics[width=120mm]{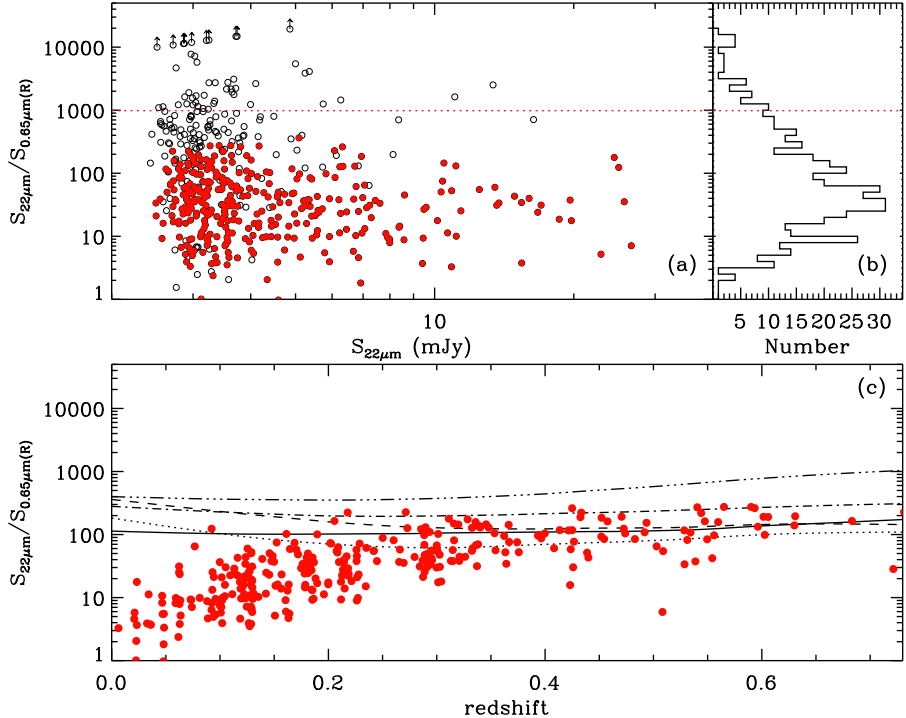}
\caption{The flux density ratio between \wise 22 $\mu$m and $R$-band
  as a function of 22 $\mu$m flux density ({\it a}), and its histogram ({\it b}).
Filled red (open black) circles are galaxies with (without) spectroscopic redshifts.
Galaxies with arrows have only an upper limit $R$-band flux density.
The flux density ratio as a function of redshift ({\it c}).
We overplot the expected ratios from several SED templates of 
  local star-forming, or AGN-host galaxies 
  \citep[M82: solid, Arp220: dotted, IRAS 22491-1808: dashed, 
  Mrk 231: dash dot, IRAS 19254-7245 South: dash dot dot dot]{pol07}.
}\label{fig-dogs}
\end{figure*}

\begin{figure*}
\center
\includegraphics[width=160mm]{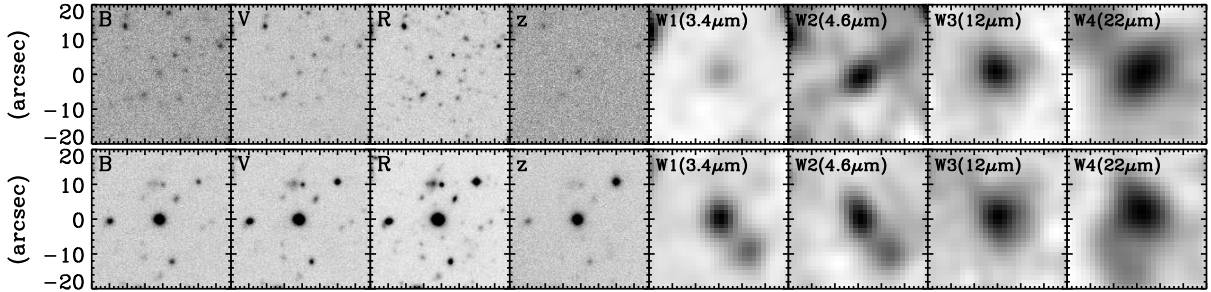}
\caption{$BVRz$ and \wise 3.4, 4.8, 12 and 22 $\mu$m 
  cutout images ($40\arcsec \times 40\arcsec$) 
  for a DOG candidate 
  (\wise ID: J092129.47+291003.1, $S_{22\mu m} = 5.00\pm0.93$ mJy, and R=23.8 mag) (Top)
  and for a typical 22 $\mu$m-selected galaxy 
  (\wise ID: J092221.93+290619.8, $S_{22\mu m} = 5.00\pm0.97$ mJy, and R=19.2 mag) (Bottom). 
}\label{fig-cut}
\end{figure*}

We plot the line ratios 
  of \wise 22 $\mu$m-selected galaxies in Figure \ref{fig-agn}(a)
  along with all SHELS galaxies (contours).
For galaxies with S/N$\geq$3 in all four lines,
  we classify them 
  (star-forming galaxies, AGNs, and composite galaxies)
  based on their relative positions 
  with respect to the demarcation lines identifying extreme starbursts \citep{kew01}
  and pure SF \citep{kau03agn}.
If the S/N of H$\beta$ or H$\alpha$ is $<3$
  (but S/Ns of [O III] and [N II] are $\geq3$),
  we adopt the 3$\sigma$ upper limit values as line fluxes
  following \citet{wes10}.
We use these upper limits to compute the line ratios,
  and classify them only if they fall 
  in the regions occupied by composite galaxies and AGNs.
There are five galaxies with very low S/N in H$\beta$;
 we show them in panel (a) with up arrows.
Similarly, if the S/N of [O III] or [N II] is $<3$
  (but S/Ns of H$\beta$ and H$\alpha$ are $\geq3$),
  we classify them only if they fall 
  in the region occupied by SF galaxies by
  adopting the 3$\sigma$ upper limits as line fluxes.
There are three galaxies with very low S/N in [O III] (down arrows).

We identify 17 broad emission-line AGNs.
The FWHM of the broad component of the Balmer or Mg II $\lambda$2800 lines
  exceeds $2000$ \kmsb.
The amplitude of the broad component exceeds the local rms of the
  continuum-subtracted spectra.

To identify additional AGNs missed by
  the spectral diagnostics,
  we use the \wise color-color diagram in panel (b).
We adopt the AGN selection criteria proposed by \citet{jar11} (solid lines in b).
There are 64 AGNs satisfying these \wise color criteria ($\sim$23\% among the galaxies in this panel).
Among them,
  34 and 5 galaxies are classified as AGN and SF based on their optical spectra, respectively;
  the five SF galaxies may contain obscured AGNs (e.g., \citealt{vei09,leejc12akari}). 
The remaining 25 galaxies cannot be classified with their optical spectra. 
We replace the galaxy classification with the one based on \wise colors
  only if their classifications based on the optical spectra are 
  either SF or undetermined (see Table \ref{tab-samp}).
  
In summary, among 259 \wise 22 $\mu$m-selected galaxies at $z<0.37$, 
  219 galaxies are classified based on the optical line ratio diagram (a):
  17\%, 36\% and 47\% as AGN, composite and SF galaxies, respectively.
If we include AGNs identified by the  \wise color-color diagram
  and by broad emission lines,
  the AGN fraction remains similar: 19\%.
For the analysis below, 
  we include composite galaxies in the sample of AGNs.
Among 78 galaxies at $z\geq0.37$,
  we have 36 AGNs ($\sim46\%$) 
  including 25 from the \wise color-color diagram and 
  11 broad-line AGNs identified from the spectroscopy.

\input{table4}

\section{Results and Discussion}\label{results}

\subsection{Dust-Obscured Galaxies (DOGs)}\label{dogs}

Among the 481 22 $\mu$m sources
  with optical counterparts in the DLS $R$-band catalog,
  144 sources do not have spectroscopic redshifts
  (see Section \ref{wise}).
We examined optical images of these 144 sources. 
The majority of these objects are indeed fainter than 
  the magnitude limit of SHELS (i.e. $R>20.6$, see Figure \ref{fig-comp}(b)).
The sources at $R<20.6$ missed by SHELS spectroscopy
  are mainly point sources or galaxies close to bright stars.
  
The optically faint 22 $\mu$m-selected galaxies
 probably result from large dust obscuration.
To study their properties in detail,
  we plot the \wise 22 $\mu$m to $R$-band flux density ratio 
  as a function of 22 $\mu$m flux density in the upper panel of Figure \ref{fig-dogs}.
We indicate the objects without optical counterparts with arrows
  based on the upper limit of the available $R$-band magnitudes (i.e. $R\sim25.2$).

Galaxies with very large $S_{24\mu m}/S_{0.65 \mu m (R)}$ are
  known as dust-obscured galaxies (DOGs, \citealt{dey08}).
\citet{dey08} used \spitzer  24 $\mu$m flux density to define DOGs 
  as objects with $S_{24\mu m}/S_{0.65 \mu m}\geq982$
  or ($R-[24]$) $\geq 14$ (Vega) mag.
They use various spectroscopic observations (primarily IR spectroscopy)
  to show that 
  most of these objects are $z\sim2$ galaxies with large ratios
  between rest-frame MIR and UV flux densities.
These large ratios seem to result from 
  abnormally large dust obscuration in the UV rather than
  from enormously large MIR emission (e.g., \citealt{pen12}).

We use the \citeauthor{dey08} criterion (i.e. $S_{22\mu m}/S_{0.65 \mu m}\geq 982$)
  to select DOGs among the 22 $\mu$m-selected SHELS galaxies
  (dotted line in the upper panel of Figure \ref{fig-dogs}).
Because of the difference in bandpasses between \spitzer 24 $\mu$m
  and \wise 22 $\mu$m,
  the \wise 22 $\mu$m flux density for DOG selection criterion
  could be smaller by $\sim10\%$ (i.e. $S_{22\mu m}/S_{0.65 \mu m}\lesssim0.9\times982$)
  if we assume the spectrum of M82 or Arp220 at $z\sim2$.
However, we retain the $S_{22\mu m}/S_{0.65 \mu m}\geq 982$
  for our selection criterion of DOGs.

We identify 48 DOG candidates in the SHELS field.
Their \wise colors suggest that 
  50\% of these objects contain AGNs  (see Figure \ref{fig-agn}(c));
  the rest of them are LIRGs or ULIRGs.
We list the DOG candidates in Table \ref{tab-dogs}
  with their $R$-band magnitudes
  and \wise flux densities at each wavelength.
Figure \ref{fig-spat} shows their spatial distribution on the sky.
We show an example of optical $BVRz$ and \wise 3.4, 4.6, 12 and 22 $\mu$m
  cutout images for a DOG candidate in the top panel of Figure \ref{fig-cut}.
We also show a typical 22 $\mu$m-selected galaxy 
  with similar 22 $\mu$m flux density in the bottom panel for comparison.
The optical counterpart for the DOG is barely visible 
  even in the very deep DLS images.

To study the redshift dependence of the flux density ratio between 
  \wise 22 $\mu$m and $R$-band,
  we plot $S_{22\mu m}/S_{0.65 \mu m}$ of galaxies
  as a function of redshift (lower panel in Figure \ref{fig-dogs}).
We overplot the flux density ratios expected from several SED templates of 
  local star-forming, or AGN-host galaxies
  \citep[M82: solid, Arp220: dotted, IRAS 22491-1808: dashed, 
  Mrk 231: dash dot, {\it IRAS} 19254-7245 South: dash dot dot dot]{pol07}.
Interestingly, all the observational data points 
  with large flux density ratios are covered by the local SED templates.
None of templates have flux density ratios larger than $\sim1000$,
  which would satisfy the DOG criterion in this redshift range.
If we compute the flux density ratios of these SED templates at $z\sim2$,
  none of these templates except IRAS 19254-7245 South\footnote{
  IRAS 19254-7245 South satisfies
  the DOG criterion at $z\sim2$, but the relevant UV SED of this galaxy is not well constrained 
  (see \citealt{berta03}).}
  satisfy the DOG criterion (see also Figure 1 in \citealt{dey08}).

\subsection{Spectral Properties of 22 $\mu$m-selected Galaxies}

\begin{figure}
\center
\includegraphics[width=85mm]{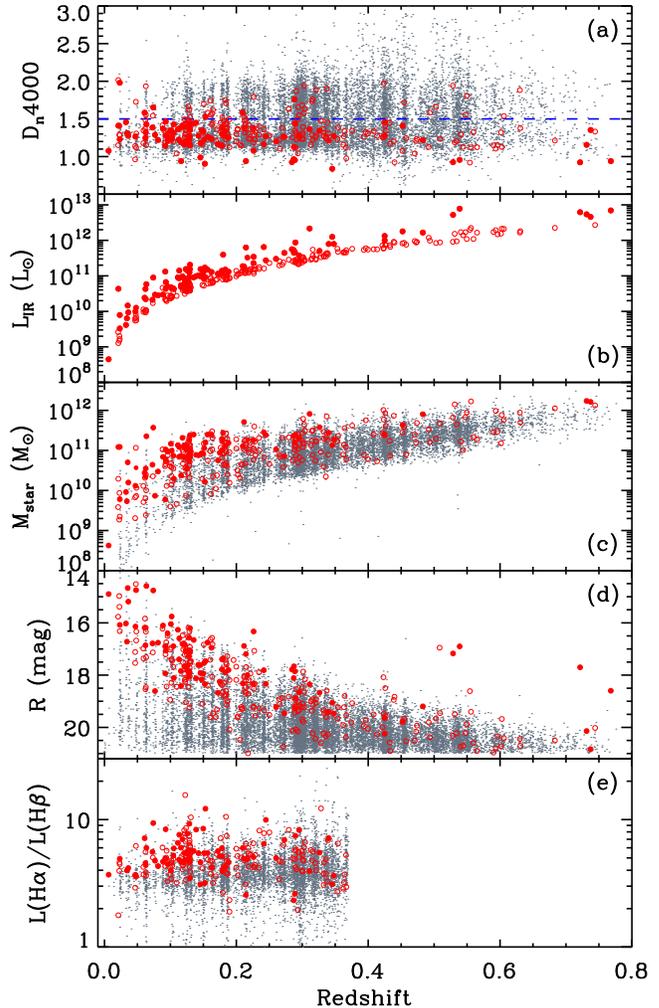}
\caption{$D_n4000$ (a), 
  IR luminosity (b), stellar mass (c),
  $R$-band magnitude (d), and
  H$\alpha$/H$\beta$ (e)
  as a function of redshift. 
Gray dots indicate all SHELS galaxies.
Red filled and open circles indicate 22 $\mu$m-selected galaxies 
  with  (S/N)$_{22\mu{\rm m}}\geq5$ and $\geq3$, respectively.
Horizontal dashed lines in (a) indicate $D_n4000=1.5$
  that divides galaxies into subsets dominated by 
  old and young stellar populations.
Red circles with $z>0.5$ and $R\approx17-18.5$ 
  are quasars from the SDSS.
}\label{fig-zdist}
\end{figure}

In this section, 
  we compare the spectral parameters of the 22 $\mu$m-selected galaxies
  with those of other SHELS galaxies.
We then discuss the properties of 
  average spectra of the 22 $\mu$m-selected galaxies.
We also show some unusual galaxies with [Ne III] strong emission.

\subsubsection{Spectral Parameters}\label{char}

In Figure \ref{fig-zdist},
  we plot several physical parameters of the 22 $\mu$m-selected galaxies
  as a function of redshift. 
In the top panel (a),
  we show the distribution of $D_n4000$ (a measure of the 4000 \AA~break).
This measure  was originally defined by \citet{bru83} as 
  $D4000$, the ratio of the average flux density $F_\nu$
  in the band $4050-4250$ \AA~ to the band $3750-3950$ \AA.
This definition was revised by \citet{bal99} as $D_n4000$
  using the narrower continuum bands 
  ($4000-4100$ and $3850-3950$ \AA).
The \citeauthor{bal99} definition is less sensitive to reddening.
This 4000 \AA~break 
  results from an accumulation of absorption lines of ionized metals in low mass stars
  at wavelength $<4000$ \AA.
The amplitude of the break is smaller in galaxies with young stellar populations
  because the opacity decreases in hot young stars.
It is larger for old metal-rich populations.
Therefore, $D_n4000$ is a useful measure of the age of the stellar population.

We adopt the $D_n4000$ measurements for SHELS galaxies
   from \citet{woods10}.
The internal systematic error in our $D_n4000$ values
  based on repeated measurements is only 4.5\%.
Our $D_n4000$ measurements from MMT and SDSS spectra 
  for the galaxies in common between SHELS and SDSS
  agree very well with a median ratio of 1.00 \citep{fab08}.
The panel (a) shows that
    the $D_n4000$ distribution for all SHELS galaxies is bimodal,
    underscoring the existence of two well-known
    galaxy populations : one dominated by old stars
    and the other with recent star formation.
Following \citet{kau03}, 
  we divide the two populations at $D_n4000=1.5$ (blue, dashed horizontal lines).
\citet{kau03} showed that galaxies with $D_n4000<1.5$
  contain young stellar populations with $\lesssim1$ Gyr.

\begin{figure*}
\center
\includegraphics[width=150mm]{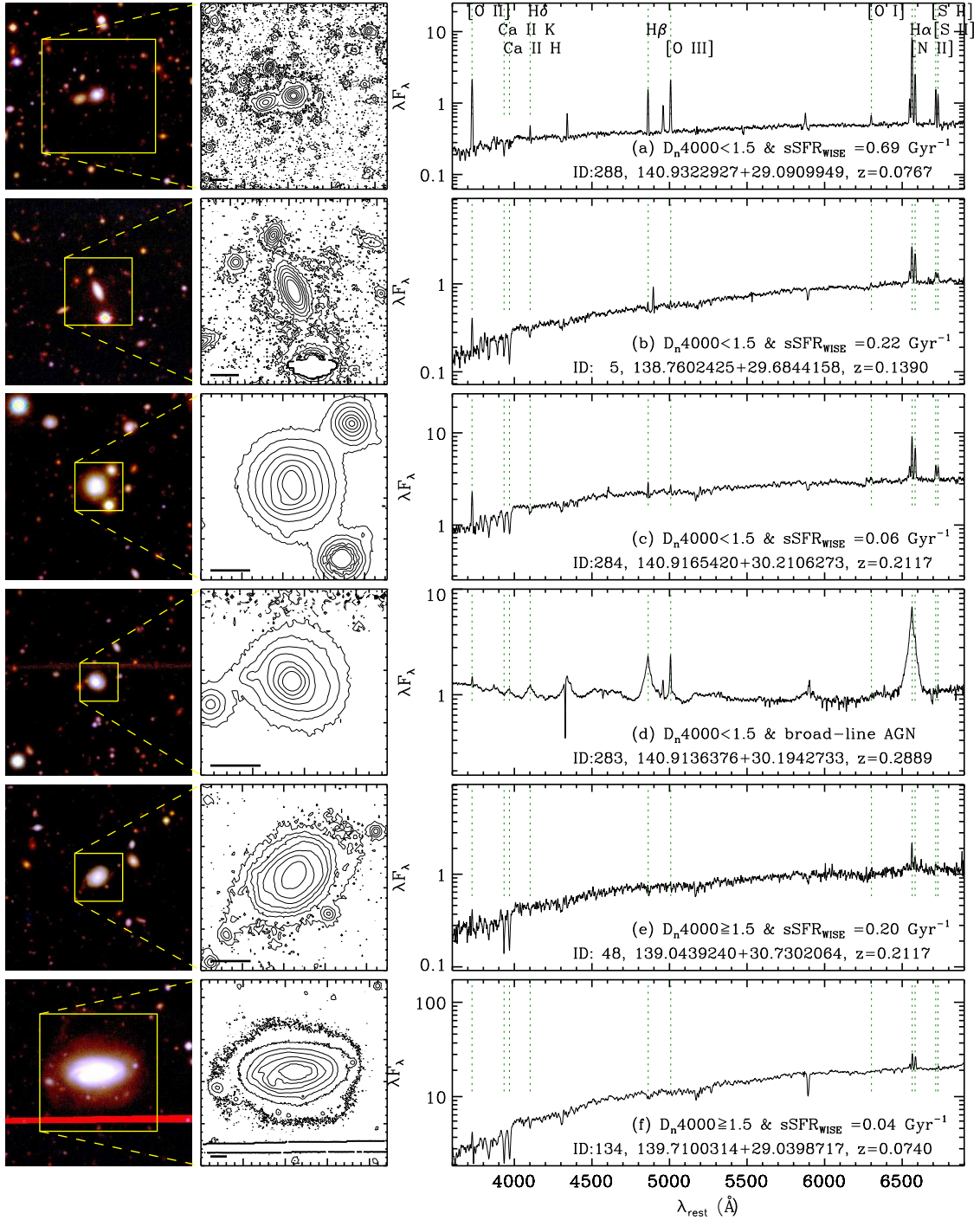}
\caption{Example MMT spectra (right panels) with
 $BVR$ color images ($90\arcsec\times90\arcsec$)
  (left panels) and 
  $R$-band contour plots (middle panels)
  for 22 $\mu$m-selected galaxies.
The north is up, and the east is to the left.
The galaxy is marked by the square indicating the size, orientation and 
  location of the contour plot in the color image.
The contour plots centered on the optical counterparts represent the intensities
  in the $R$-band image. 
The size of each contour plot is 40 kpc$\times40$ kpc.
The thick, horizontal bar represents 5 arcsec in each contour plot. 
The optical spectra are in unit of $10^{-17}$ erg s$^{-1}$ cm$^2$.
ID, SHELS ID and redshift (see Table \ref{tab-samp}) of each source are shown.
}\label{fig-sed}
\end{figure*}

If 22 $\mu$m-selected galaxies are 
  powered by recent SF or nuclear activity,
  their stellar populations should be young (i.e. small $D_{n}$4000).
As expected,
  most 22 $\mu$m-selected galaxies have small $D_{n}$4000 
  ($\sim86\%$; $60\%$ of all SHELS galaxies have $D_{n}4000<1.5$), 
  but there are some 22 $\mu$m-selected galaxies with 
  large $D_{n}$4000 (see Section \ref{starb}).
We use the Kolmogorov-Smirnov (K-S) test to determine whether 
  the $D_{n}$4000 distributions of 22 $\mu$m-selected and all SHELS galaxies
  are drawn from the same distribution.
To have a fair comparison,
  we construct 1000 trial data sets 
  by randomly selecting galaxies among all SHELS galaxies
  to have the same number of galaxies and
  to have the same redshift distribution as the 22 $\mu$m-selected galaxies.
We run the K-S test between these trial data sets and 
  the 22 $\mu$m-selected galaxies.  
For nearly all cases ($\sim99.8\%$),
  the K-S test rejects
  the hypothesis that the two distributions are extracted
  from the same parent population with a significance level of $>99.7\%$,
  confirming that 22 $\mu$m-selected galaxies are 
  inconsistent with a random subsample of SHELS galaxies.

\begin{figure*}
\center
\includegraphics[width=110mm]{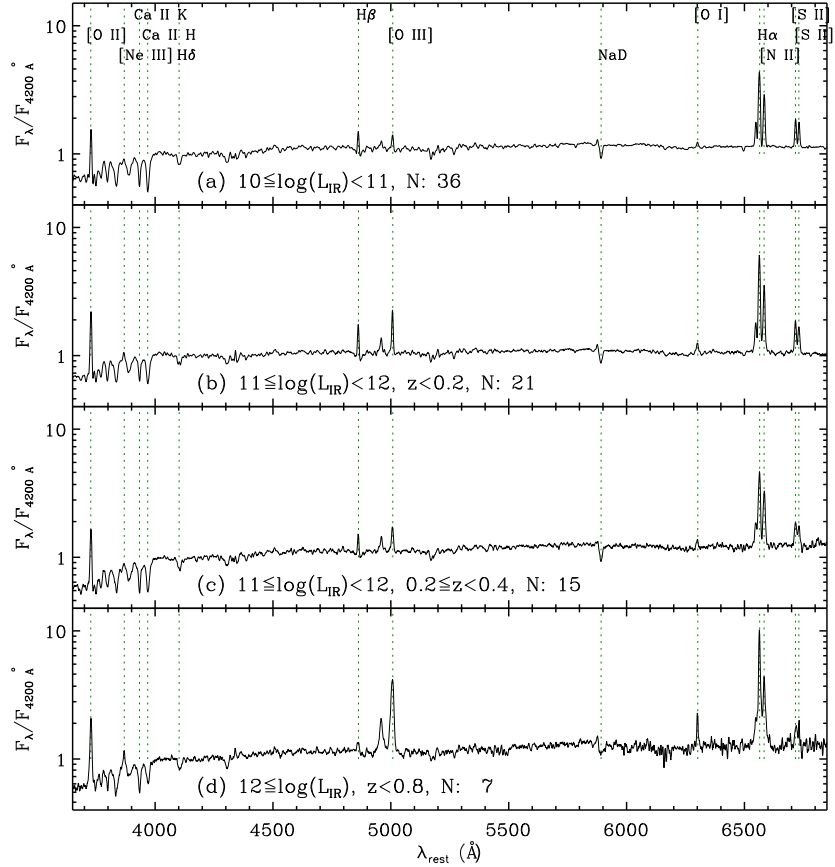}
\caption{Stacked optical spectra
  of 22 $\mu$m-selected galaxies
  according to their IR luminosities and redshifts.
N is the number of spectra in the stack.
}\label{fig-stack}
\end{figure*}

Panel (b) shows IR luminosities of 22 $\mu$m-selected galaxies
  as a function of redshift.
This figure clearly shows that the detection limit increases with redshift.
LIRGs ($L_{IR}\ge 10^{11}$ $L_\odot$) begin to appear at $z\approx0.09$, and
  ULIRGs ($L_{IR}\ge 10^{12}$ $L_\odot$) begin to appear at $z\approx0.31$.
At fixed redshift,
  the stellar masses of 22 $\mu$m-selected galaxies 
  are larger than those of 22 $\mu$m-undetected galaxies (panel c).
Similarly, the apparent $R$-band magnitudes
  of 22 $\mu$m-selected galaxies are smaller than
  than those of 22 $\mu$m-undetected galaxies (panel d).
Because there is a close connection between the stellar masses and
  the SFRs (i.e. IR luminosities)
  (e.g., \citealt{noe07sf,elb07}),
  the larger masses (or small $R$-band magnitudes) 
  of 22 $\mu$m-selected galaxies
  result primarily from the IR detection limits in panel (b).

Panel (e) shows the flux ratio between $H\alpha$ and $H\beta$ (i.e. Balmer decrement)
  as a function of redshift.
As expected, 
  most 22 $\mu$m-selected galaxies show $H\alpha/H\beta$ values 
  larger than the intrinsic ratio in the nominal case B recombination
  for $T=10,000$ K with no dust (i.e. $H\alpha/H\beta\sim$2.87, \citealt{ost06}). 
In the redshift range where both $H\alpha$ and $H\beta$ are measured (i.e. $z<0.37$),
  the ratio does not change with redshift.

We show a sample of $BVR$ color images,
  $R$-band contour plots, and optical spectra
  for 22 $\mu$m-selected galaxies
  with various sSFRs
  and $D_n4000$ values in Figure \ref{fig-sed}.
The figure shows a variety of spectral features depending on
  their sSFRs and $D_n4000$.
As expected, all of them have H$\alpha$ or [O II] emission lines,
  indicating their SF or nuclear activity.
The other emission lines including H$\beta$ and [O III]
  are also strong for the galaxies with small $D_n4000$ (top four panels),
  but are very weak for those with large $D_n4000$ (bottom two panels).
The Ca II H and K absorption lines,
  indicative of old stellar populations,
  are weak in the spectrum of the galaxy with the largest sSFR (a).
These lines are not visible in the spectrum of the broad-line AGN (d).
However, these lines are strong for the galaxies 
  with large $D_n4000$ (bottom two panels).

\begin{figure}
\center
\includegraphics[width=80mm]{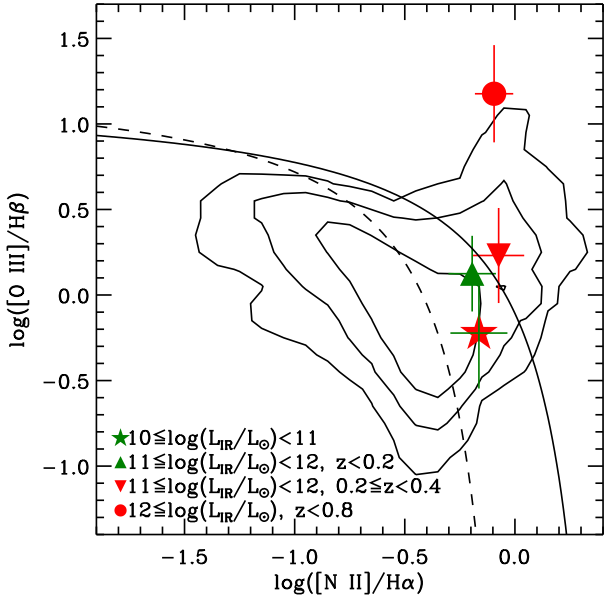}
\caption{The line ratio diagram 
  for the stacked spectra in Fig. \ref{fig-stack}.
The solid and dashed lines in (a) indicate the extreme starbursts \citep{kew01}
  and pure SF limits \citep{kau03agn}, respectively.
The contours indicate the distribution of all SHELS galaxies at $z<0.37$
  adopted from Fig. \ref{fig-agn}a.
}\label{fig-stackagn}
\end{figure}

\begin{figure*}
\center
\includegraphics[width=150mm]{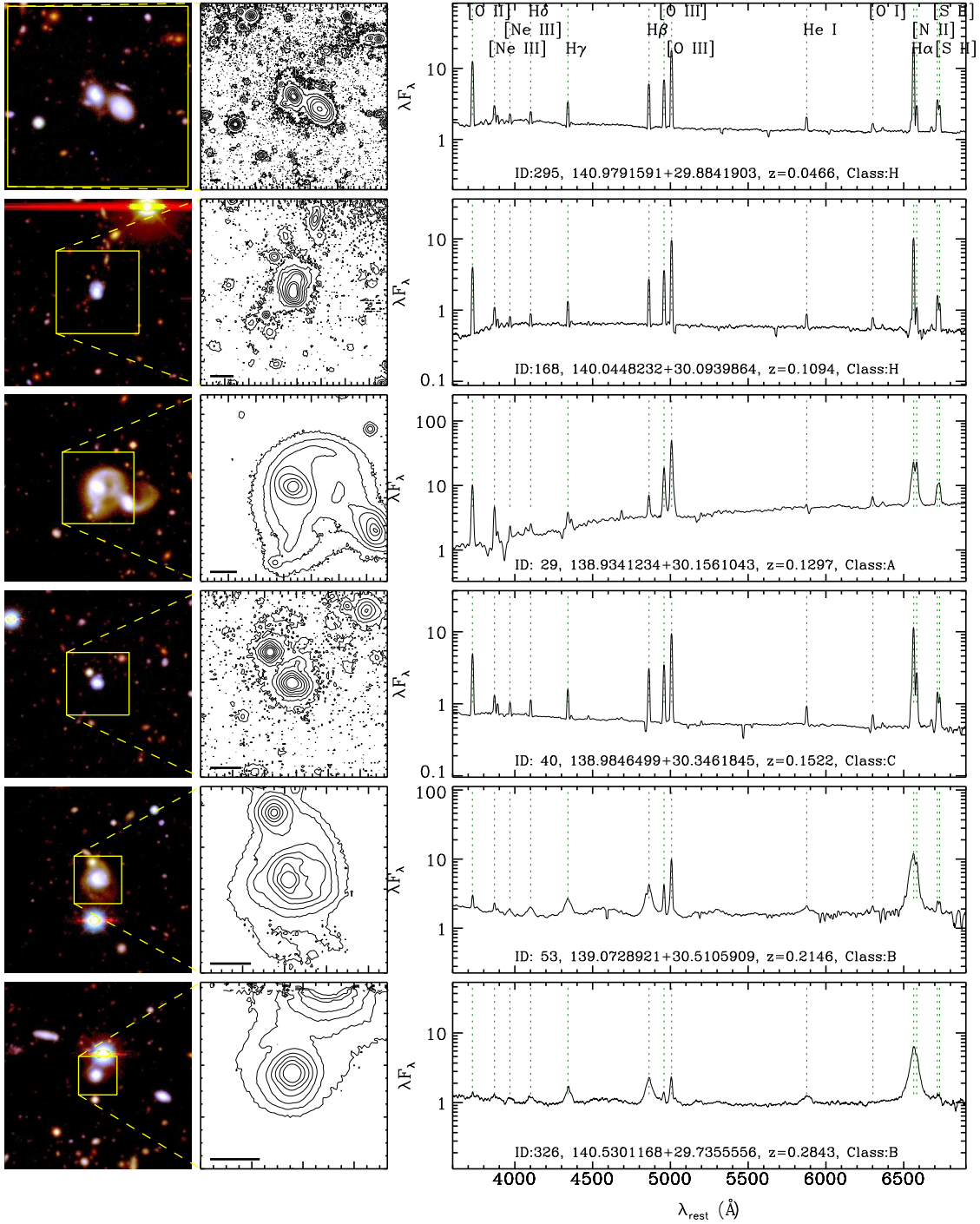}
\caption{Same as Fig. \ref{fig-sed}, but for [Ne III] strong galaxies.
For each source,
  we show ID, SHELS ID, redshift and SF/AGN classification in Table \ref{tab-samp} 
  (Class: H (SF), C (Composite), 
  A (AGN from optical spectra), M (AGN from MIR colors), B (broad-line AGN)).
}\label{fig-emsed}
\end{figure*}

\begin{figure*}
\center
\includegraphics[width=150mm]{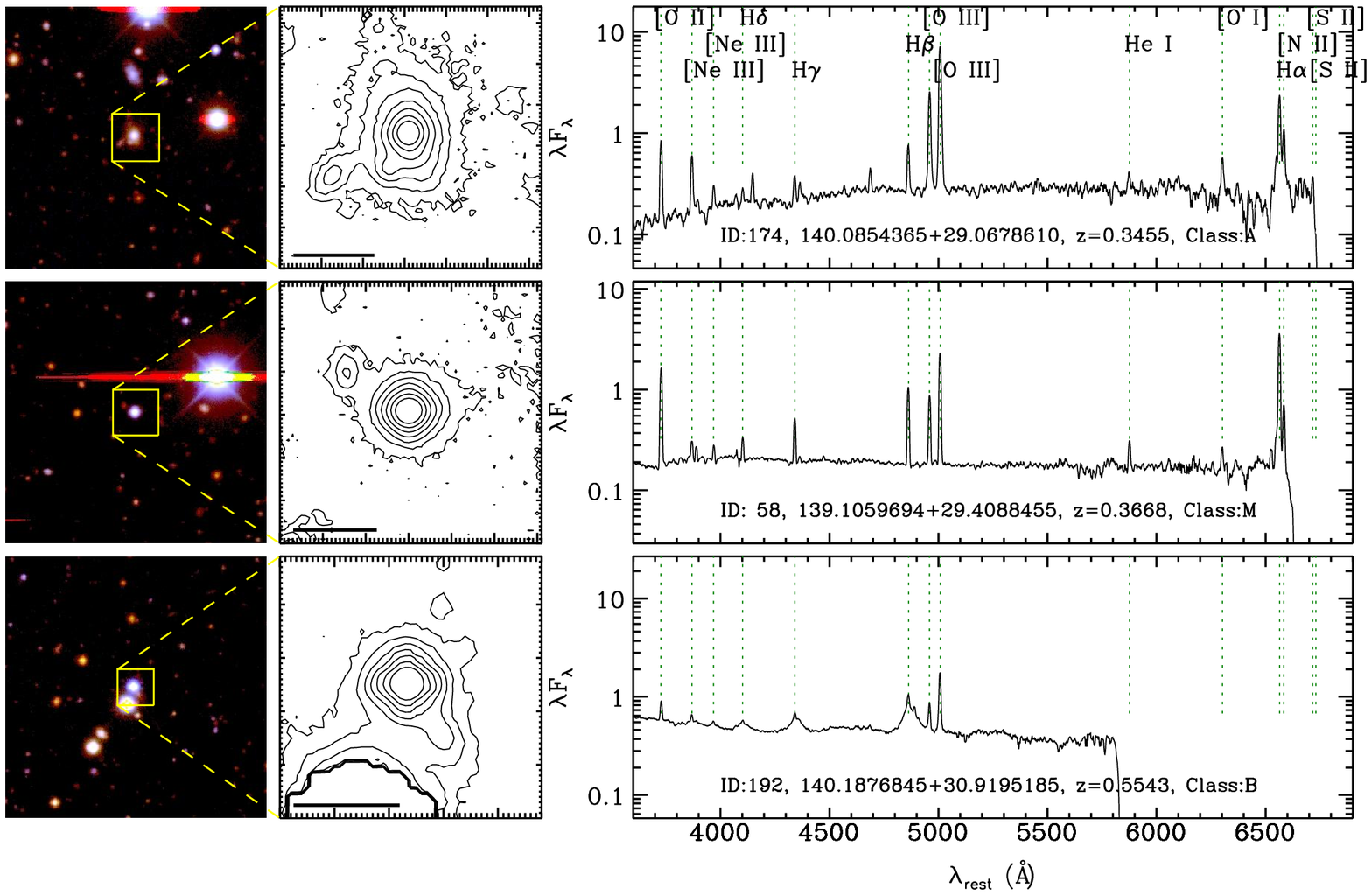}
\center
Fig. \ref{fig-emsed}.--- Continued
\end{figure*}

\subsubsection{Average Spectra}\label{stack}

To examine the typical optical spectral features for 22 $\mu$m-selected galaxies,
  we show the rest-frame median-stacked spectra of these galaxies
  binned in IR luminosity (Figure \ref{fig-stack}).
Because LIRGs are distributed over a wide redshift range, 
  we divide them into two subsamples
  based on their redshifts (i.e. $z<0.2$ and $0.2\leq z<0.4$).
We exclude the broad-line AGNs.
Thanks to the wide wavelength coverage of Hectospec spectra,
  several spectral features including $D_n4000$ and the H$\alpha$ emission line
  are well represented for all subsamples.
For example, many emission lines including [O II], [O III] and H$\alpha$
  are clearly visible, indicating SF and/or nuclear activity.
Higher-order Balmer absorption lines at $<3900$ \AA~are also prominent 
  for all subsamples.
These lines are characteristic of short-lived ($\lesssim 1$Gyr) A-type stars,
  typically present in post-starburst galaxies
  that may have experienced starbursts within the last Gyr but
  with no current SF \citep{dre83}.
However, we do not find any galaxies in our sample satisfying
  the selection criteria for post-starburst galaxies
  (e.g., strong H$\delta$ absorption line and no [O II] and H$\alpha$ emission lines),
  probably because our samples are 22 $\mu$m-selected galaxies
  that are still forming stars.

Some absorption lines including Na D and Ca II H and K,
  indicative of old stellar populations, are also visible in Figure \ref{fig-stack}
  (see \citealt{cap08,cap09} for stacked spectra of \spitzer 24 $\mu$m-selected
  galaxies at similar redshift for comparison).
The comparison of stacked spectra among subsamples is interesting.
As expected, H$\alpha$ and [O II] emission lines (SFR indicators) 
  strengthen with increasing IR luminosity (from top to bottom).
The strength of the [O III] line relative to H$\beta$ also grows with increasing IR luminosity,
  which can suggest an increase in nuclear activity.

To better demonstrate this trend of nuclear activity with IR luminosity,
  we plot the emission line ratios of
  [O III]/H$\beta$ and [N II]/H$\alpha$ 
  for these stacked spectra in Figure \ref{fig-stackagn}.
While the [NII]/H$\alpha$ ratios do not change much with IR luminosity,
  the [O III]/H$\beta$ ratios increase significantly
  with increasing IR luminosity.
Although there are two LIRG samples at $z<0.2$ and at $0.2\leq z<0.4$,
    the mean IR luminosity for LIRGs at $0.2\leq z<0.4$ 
    slightly exceeds the mean IR luminosity of LIRGs at $z<0.2$ 
    ($\langle{\rm log(L_{IR})}\rangle=11.6$ vs. $=11.1$)
    as a result of selection toward
    increasing luminosity limit with redshift (see Figure \ref{fig-zdist}(b)). 
The number fraction of AGN-host galaxies 
  in each subsample also increases with IR luminosity 
  from 44\% [10$\leq$log(L$_{\rm IR}$)$<$11] to 86\% [12$\leq$log(L$_{\rm IR}$)],
  consistent with previous results (e.g., \citealt{vei95,vei02,yuan10,hwa10lirg}).

\subsubsection{[Ne III] Strong Galaxies}\label{neon}
  
During the visual inspection of 
  all of the spectra of the 22 $\mu$m-selected galaxies,
  we identified some unusual galaxies with strong emission lines
  not used for the line ratio diagram in Figure \ref{fig-agn}.
These lines include [Ne III] 3869, 3968 \AA~(and sometimes He I 5876 \AA) that 
  require hard ionizing radiation such as AGN or 
  extremely young massive stars \citep{ost06}.
To select these galaxies in an objective way,
  we use the equivalent widths (EWs) of [Ne III] 3869, 3968 \AA~lines:
  EW$_{\rm [Ne III] 3869}<-55$ \AA~and EW$_{\rm [Ne III] 3968}<-45$ \AA.
Among the 317 22 $\mu$m-selected galaxies at $z<0.8$,
  we find nine galaxies ($\sim2.8\%$).
This fraction is similar to the fractions of [Ne III] $\lambda$3869, 3968 strong galaxies
  in SDSS \citep[$1.2-2.7\%$]{sb12}.
Reasonable changes in the selection criteria do not change the sample.
Figure \ref{fig-emsed} shows the optical spectra along with optical color images
  and IDs (see Table \ref{tab-samp}).
  
\begin{figure*}
\center
\includegraphics[width=160mm]{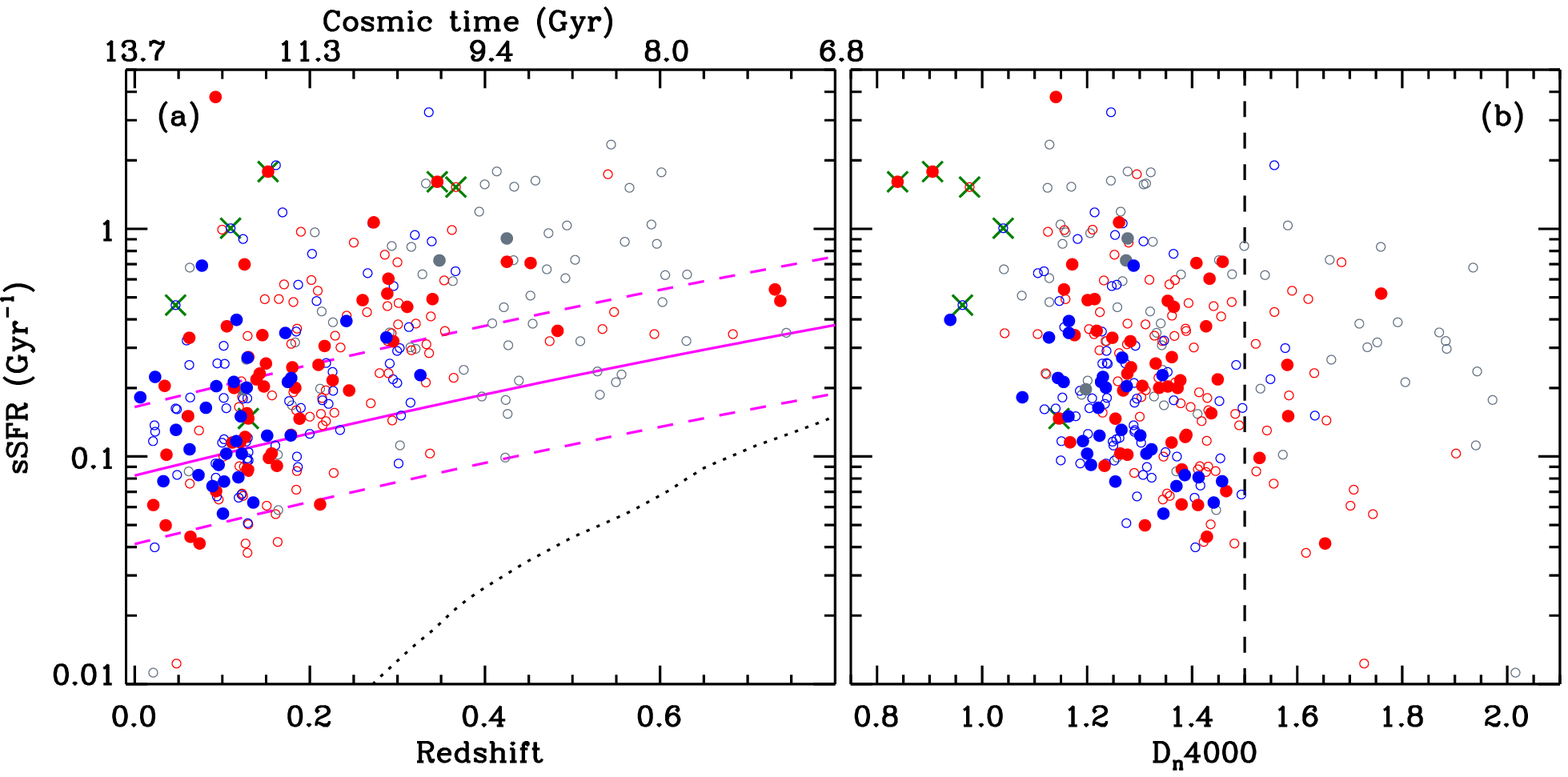}
\caption{Redshift evolution of sSFRs of 
  \wise 22 $\mu$m-selected galaxies ({\it a}).
Filled and open symbols are
 galaxies with S/N$_{22\mu{\rm m}}\geq 5$ and 
 S/N$_{22\mu{\rm m}}\geq 3$, respectively.
Red and blue colors indicate AGN-host and SF galaxies, respectively, and
  gray colors are those without SF/AGN classification.
The crosses are [Ne III] strong galaxies (see Section \ref{neon}).
The pink solid line is the median trend of normal SF galaxies in \citet{elb11}.
Two dashed lines indicate the boundaries for starburst and quiescent galaxies.
The dotted line is the detection limit for galaxies with 
  $M_{\rm star}\approx3\times10^{12}$ $M_\odot$.
 ({\it b}) sSFRs vs. $D_n4000$ for \wise 22 $\mu$m-selected galaxies.
Symbols are the same as in (a).
}\label{fig-zsfr}
\end{figure*}

The [Ne III] line is more conspicuous in AGNs than in star-forming galaxies \citep{rola97}.
Actually, seven out of nine galaxies in our sample host AGNs.
However, the hard radiation for [Ne III] line can also be provided by 
  very young stars in HII regions \citep{shi90}.
There are two galaxies with SF class (see Figure \ref{fig-emsed}).
Moreover, many [Ne III] strong galaxies have large sSFRs (see Figure \ref{fig-zsfr}), 
  suggesting that [Ne III] lines originate from SF as well as AGNs.

\begin{figure}
\center
\includegraphics[width=80mm]{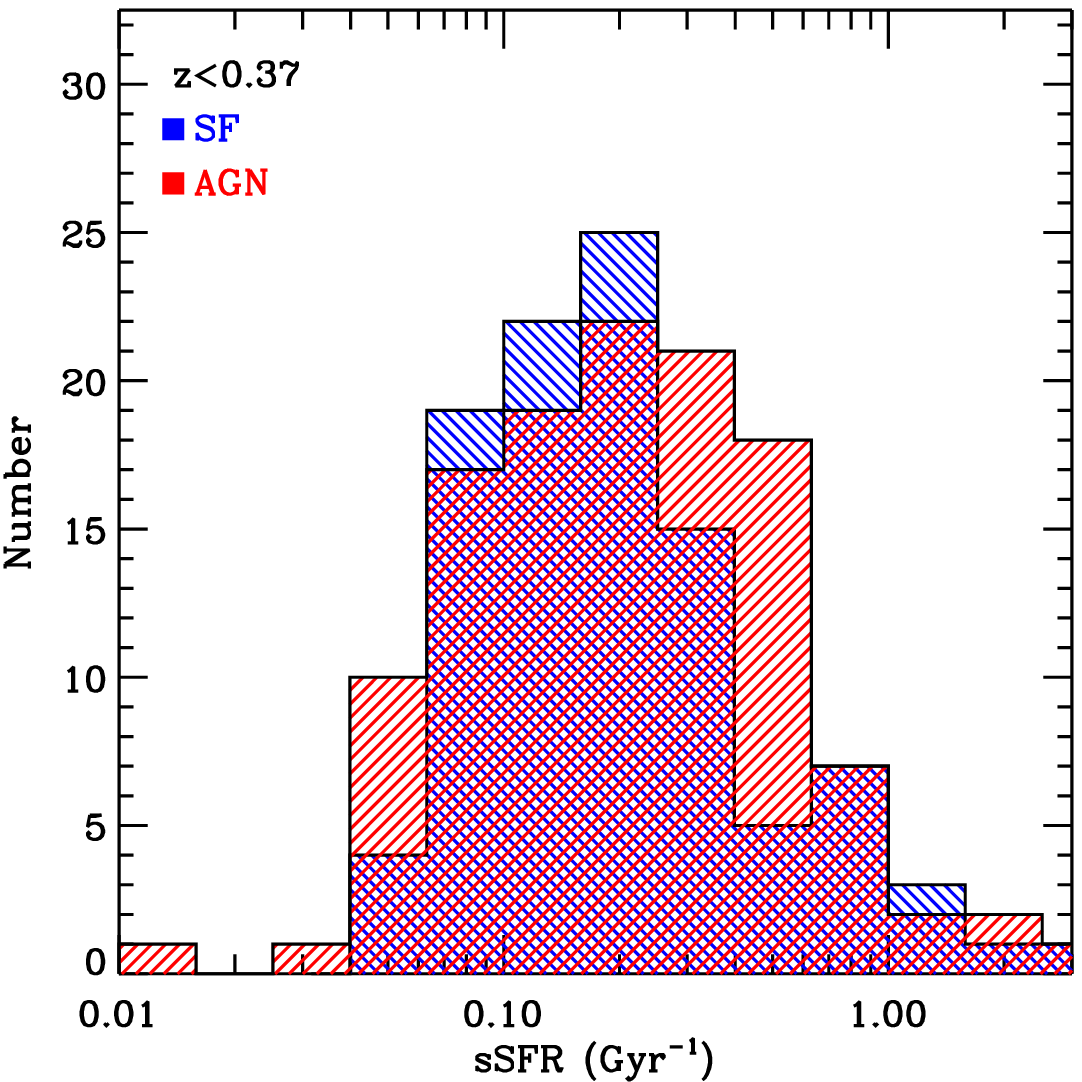}
\caption{Distribution of sSFRs
  of the 22 $\mu$m-selected galaxies with S/N$_{22\mu{\rm m}}\geq 3$
  depending on SF/AGN classification.
Galaxies with AGN and SF classes at $z<0.37$
 are denoted by hatched histograms with
  orientation of 45$^\circ$ ($//$ with red color) and 
  of 315$^\circ$ ($\setminus\setminus$ with blue color) 
  relative to horizontal, respectively.
}\label{fig-starb}
\end{figure}

\subsection{Correlation between specific SFRs and Optical Spectral Properties}\label{starb}

To study how the SFA of 22 $\mu$m-selected galaxies
  evolves with cosmic time,
  we plot their sSFRs
  as a function of redshift in the left panel of Figure \ref{fig-zsfr}.
The sSFRs of 22 $\mu$m-selected galaxies increase on average with redshift.
We overplot the detection limit for galaxies with 
  the largest stellar mass in our sample 
  ($M_{\rm star}\approx3\times10^{12}$ $M_\odot$) as a dotted line.
The sSFRs of 22 $\mu$m-selected galaxies are much larger than
  than the limit,
  suggesting that the increase of sSFRs with redshift is not strongly affected by
  the detection limit.
    
We overplot the mean evolutionary trend of IR-selected star-forming galaxies
  suggested by \citet[solid line]{elb11}.
The \wise 22 $\mu$m-selected galaxies, on average, follow
  this evolutionary trend
  with some outliers with large sSFRs. 
\citeauthor{elb11} introduced a ``starburstness'' parameter
  that is the ratio of sSFR of a galaxy to the median sSFR at a given redshift.
Because this parameter corrects the underlying trend of sSFR with redshift,
  it is useful in making a fair comparison of the SFA of galaxies 
  in a wide redshift range (e.g., $0<z<3$). 
However, the redshift range for our sample is small 
  (most of galaxies relevant to the following analysis are at $z<0.37$), and 
  the slope of the evolutionary trend can differ
  depending on the sample selection 
  (see \citealt{kar11} and \citealt{elb11} for detailed discussion 
  on this issue including the slope of SFR-$M_{\rm star}$ relation).
We thus simply use sSFR as a proxy for the SFA of galaxies for the following analysis.
If we use the starburstness rather than sSFR,
  the results do not change.
  
Because H$\alpha$ bright galaxies are all detected at 22 $\mu$m,
  the general trend in Figure \ref{fig-zsfr} does not change
  if we use H$\alpha$ luminosity rather than IR luminosity 
  (see \citealt{wes10} for the analysis based on H$\alpha$ luminosity).
However, the H$\alpha$ fluxes
  can be easily contaminated by the presence of AGN,
  they are often uncertain because of large extinction corrections for dusty galaxies, and
  they can not be measured for high-z galaxies because of 
    the limited spectral coverage.
The use of 22 $\mu$m flux densities thus provides a 
 more extensive view of SFA of galaxies.

The right panel of Figure \ref{fig-zsfr} shows 
  the sSFRs as a function of $D_n4000$.
As expected,
  the sSFRs of galaxies with small $D_n4000$ (e.g., $<1.2$),
  on average, 
  are larger than those with large $D_n4000$ (e.g., $>1.2$).
[Ne III] strong galaxies (crosses) in general have large sSFRs (e.g., $\gtrsim0.4$ Gyr$^{-1}$),
  indicating their intense SFA.

\begin{figure*}
\center
\includegraphics[width=140mm]{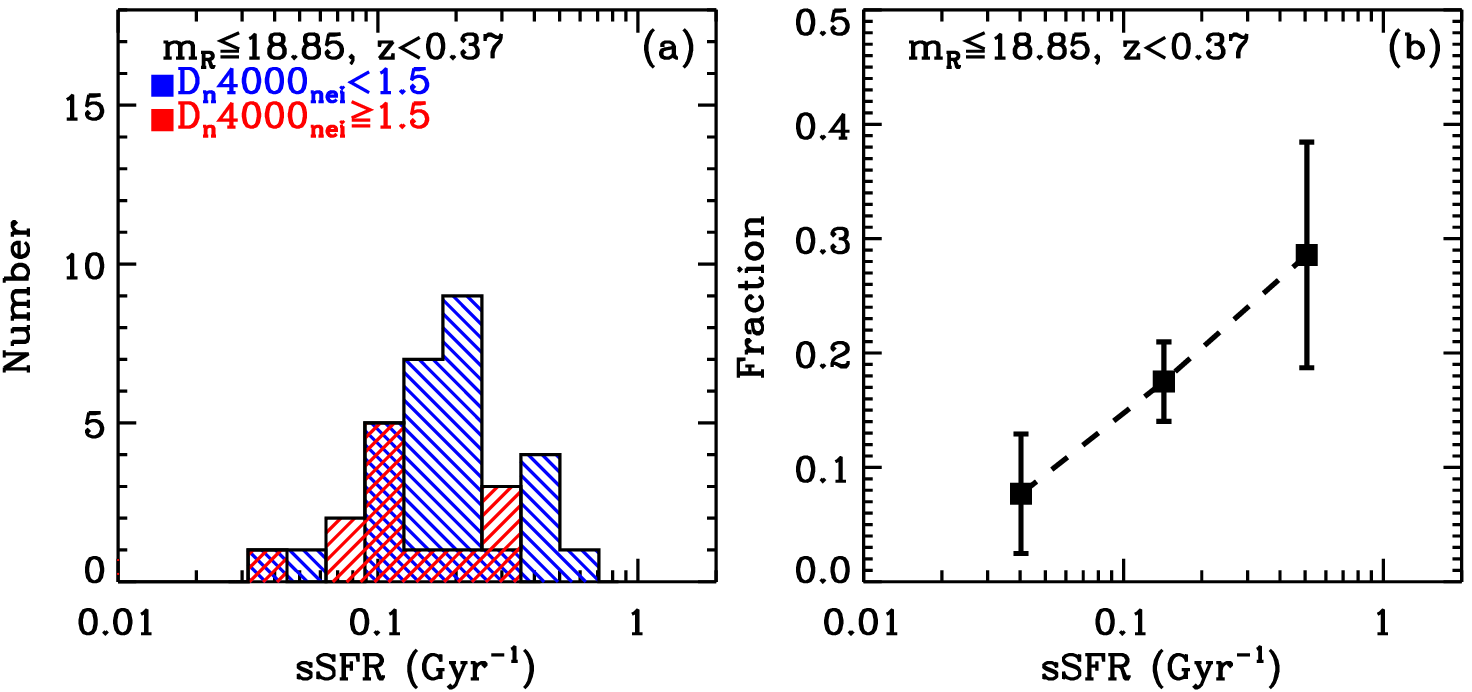}
\caption{The sSFR distribution for
  the 22 $\mu$m-selected galaxies 
  with S/N$_{22\mu{\rm m}}\geq 3$
  depending on $D_n4000$ of the nearest neighbor galaxy
  (i.e. $D_n4000_{\rm nei}$) ({\it a}).
Galaxies with $D_n4000_{\rm nei}\geq1.5$ and $D_n4000_{\rm nei}<1.5$
   are denoted by hatched histograms with
  orientation of 45$^\circ$ ($//$ with red color) and 
  of 315$^\circ$ ($\setminus\setminus$ with blue color) 
  relative to horizontal, respectively.
The fraction of galaxies with close neighbors
  with $D_n4000_{\rm nei}<1.5$
  among 22 $\mu$m-selected (S/N$_{22\mu{\rm m}}\geq 3$) galaxies
  as a function of sSFR ({\it b}).
The associated errorbars indicate Poisson uncertainties.
}\label{fig-inter}
\end{figure*}

There are some 22 $\mu$m-selected galaxies
  with large $D_n4000$ (e.g., $>1.5$),
  suggesting star formation in a system dominated by 
  an old stellar population.
Many of these objects are indeed morphologically early-type 
  (with signs of disturbances in some cases).
They show some emission lines in their optical spectra, 
  indicating SF and/or nuclear activity 
  (see Figure \ref{fig-sed}(e-f)).
These galaxies are similar to 
  star-forming (or blue), early-type galaxies
  found both in low-$z$ and in high-$z$ universe 
  (e.g., \citealt{fuk04,jhlee06,jhlee10beg,hwa12a2199});
  SF or nuclear activity probably results from
  recent galaxy interactions or mergers.
We show the distribution
  of these objects in the optical line ratio (a) and \wise color-color (b) diagrams
  of Figure \ref{fig-agn}.
There are a small number of galaxies satisfying the MIR AGN selection criteria.
However, the \wise colors for the majority of these 22 $\mu$m-selected galaxies
  with large $D_n4000$ (see panel b)
  are consistent with those of normal spiral galaxies:
  two galaxies are in the color range of elliptical galaxies.
Among 38 22 $\mu$m-selected galaxies with $D_n4000>1.5$ at $z<0.37$,
  we can classify 24 galaxies based on the optical line ratios (panel a):
  46\%, 37\% and 17\% as AGN, composite and SF galaxies, respectively.
The large fraction of AGN-host galaxies among the emission-line early-type galaxies
  is consistent with the results based on low-$z$ galaxies (e.g., \citealt{sch07, jhlee08}).
This large fraction can suggest that AGNs play a role in the evolution of 
  emission-line early-type galaxies \citep{sch07}.

\subsubsection{AGN vs. Star-forming Galaxies}

The 22 $\mu$m-selected galaxies are mostly SF galaxies,
  but they can also harbor AGNs.
Here, we investigate the SFA of the 22 $\mu$m-selected galaxies
  containing AGNs (see Section \ref{agnsel} for the description of AGN selection).
  
One interesting feature in the right panel of Figure \ref{fig-zsfr} is
  that the sSFR distribution of AGN-host galaxies (red symbols)
  is not distinguishable from purely SF galaxies (blue symbols).
To better demonstrate this coincidence, we compare the sSFR distribution of AGNs with 
  that of SF galaxies in Figure \ref{fig-starb}.
We use the galaxies (with S/N$_{22\mu{\rm m}}\geq 3$) at $z<0.37$ where 
  we can classify them based on the optical spectra.
The distribution of the two samples does not differ significantly.
The K-S test rejects
  the hypothesis that the sSFR distribution of AGN and SF galaxies
  are extracted
  from the same parent population with only a significance level of $83\%$,
  suggesting the similar underlying distribution of the two subsamples.

This result is consistent with the recent results in \citet{mul12},
   who showed that the majority of moderate luminosity X-ray AGNs 
   at $0.5 < z < 3$ reside in normal SF galaxies
   (see also \citealt{san12} and \citealt{rov12} for the most luminous X-ray AGNs).
These results suggest that the nuclear activity in dusty star-forming galaxies
  at these redshifts
  is mainly fueled by internal processes (i.e. normal SF) 
  rather than by violent mergers (i.e. starburst).
Morphological analysis of AGN-host galaxies at high redshift
  also provides little evidence of recent merger events (e.g., \citealt{sch11, koc12})
  in contrast with AGN-host galaxies in the local universe 
  (e.g., \citealt{ell11,hwa12agn}).

\subsubsection{Role of the Nearest Neighbor Galaxy in Changing sSFR}

The SF or nuclear activity of galaxies
  is strongly affected by interactions
  with nearest neighbor galaxies
  (e.g., \citealt{bar00,woods06,gel06,ell08,ell11,pc09,hwa12agn}).
To examine the role of neighbors in changing the SFA of galaxies,
  we show the sSFR distribution of the 22 $\mu$m-selected galaxies
  depending on $D_n4000$ of their neighbors (i.e. $D_n4000_{\rm nei}$)    
  in the left panel of Figure \ref{fig-inter}.

To find the galaxies in potentially interacting pairs,
  we search for the nearest neighbor galaxy among galaxies
  with magnitudes brighter than $M_R=M_{R,\rm target}+\Delta M_R$ and 
  with relative velocities less than 
  $\Delta \upsilon/(1+z)=|\upsilon_{\rm neighbors}-\upsilon_{\rm target}|/(1+z)=500$ km s$^{-1}$.
Following \citet{woods10},
  we adopt $\Delta M_R=1.75$,
  corresponding to a luminosity (or mass) ratio $>1/5$.
This criterion selects the galaxies in major interacting pairs;
  these objects should be the most effective 
  in triggering SFA \citep{woods07,cox08,hwa10lirg}.  
To have a fair sample of neighbor galaxies in our sample,
   we select the target galaxies
   among those with $m_R<(20.6-1.75)$
   where $m_R=20.6$ is the magnitude limit of SHELS.
Because of this strict magnitude limit,
  the redshift range for the galaxies in this analysis
  is small ($z<0.37$, see Figure \ref{fig-zdist}d).
The redshift effect on the change of sSFR is thus minimized.

We use $\Delta \upsilon/(1+z)=500$ km s$^{-1}$,
  the typical maximum velocity difference used for exploring 
  close pairs in the local universe
  (see Fig. 2 of \citealt{bar00} and Fig. 1 of \citealt{park08}).   
Because we are interested in the effect of neighbors interacting with target galaxies,
  we restrict our sample to close galaxy pairs
  with a projected spatial separation of $<150$ kpc.
This separation is much smaller than the typical virial radius of late-type galaxies
  where galaxy properties begin to change because of galaxy interactions
  (e.g., $r_{vir}\approx 340$ $h_{70}^{-1}$kpc with $M_r=-20$, \citealt{park08}).
There are 47 22 $\mu$m-selected galaxies 
  at $z<0.37$ with close neighbors for the following analysis.

Panel (a) of Figure \ref{fig-inter} shows that
  the peak sSFRs for galaxies 
  with $D_n4000_{\rm nei}<1.5$ and $>1.5$ are different.
The median sSFR for galaxies 
  with $D_n4000_{\rm nei}\geq1.5$ and $D_n4000_{\rm nei}<1.5$
  is $0.12\pm0.08$ and $0.17\pm0.09$, respectively.
The K-S test rejects the hypothesis that
  two distributions are extracted
  from the same parent population with a significance level of $97\%$.
Although the statistical significance is not very high,
  this result can be consistent with the results that
  interactions between gas-rich galaxies (i.e. $D_n4000_{\rm nei}<1.5$)
  increase the SFRs;
  interactions involving gas-poor galaxies (i.e. $D_n4000_{\rm nei}\geq1.5$)
  do not (e.g., \citealt{woods07,pc09,xu10}).
This trend is also seen even
  for FIR-selected galaxies
  both in low-$z$ and in high-$z$ universe \citep{hwa10lirg,hwa11inter}.  

Figure \ref{fig-inter}(b)
  shows the fraction of 22 $\mu$m-selected galaxies 
  with projected close neighbors with $D_n4000_{\rm nei}<1.5$
  as a function of sSFR.
The fraction of galaxies with close neighbors
  clearly increases with sSFR.
This dependence strongly supports the idea that
  the sSFRs of galaxies with close neighbors in this redshift range ($z<0.37$) 
  are indeed enhanced
  by the interactions with gas-rich ($D_n4000<1.5$) galaxies 
  (e.g., \citealt{woods07,woods10,hwa11inter}).

Among the six [Ne III] strong galaxies in this redshift range,
  only one galaxy has a close neighbor (ID 29).
If the large SFRs of these objects 
  result from very recent galaxy interactions or mergers,
  this result suggests that they are in the late stage of merging;
  when two galaxies merge, the new nearest neighbor galaxy
  of the merger product may be far away.
The optical images for some of these objects show
  signs of tidal disturbances (see Figure \ref{fig-emsed}),
  supporting this suggestion that they are merger remnants.
An extensive analysis of the morphology of these objects
  along with nearby analogs (e.g., \citealt{sb12})
  and a larger sample would clarify this issue.

\section{Conclusions}\label{sum}

Using a dense redshift survey (SHELS) covering a 4 square degree region of
   a deep imaging survey (DLS),   
  we obtain a nearly complete identification of optical counterparts of 
  {\it WISE} 22 $\mu$m sources.
The properties of the {\it WISE}-selected sample are:

\begin{enumerate}

\item Among 507 \wise 22 $\mu$m sources 
  with (S/N)$_{22\mu{\rm m}}\geq3$ ($\approx S_{22\mu m}\gtrsim2.5$ mJy) 
  in the SHELS field,
  we identify 481 sources with optical counterparts 
  in the very deep, DLS $R$-band source catalog (down to $R<25.2$).
There are 337 galaxies at $R<21$ with spectroscopic data.

\item The 22 $\mu$m-selected galaxies are
  dusty star-forming galaxies
  with young stellar populations (i.e. small $D_n4000$).
Most of them are at $z<0.8$ with a median redshift of
  $\langle z\rangle=0.2$.
Their IR luminosities are in the range
 $4.5\times10^8  ({\rm L}_\odot) \lesssim L_{IR} \lesssim 7.8\times10^{12}  ({\rm L}_\odot)$
 with a median $L_{IR}$ of $1.3\times10^{11}$ (${\rm L}_\odot$).

\item There are some ($\sim$14\%) 22 $\mu$m-selected galaxies
  with large $D_n4000$ ($>1.5$),
  indicating star formation and/or nuclear activity in a system dominated by 
  an old stellar population.
Many of these objects with emission lines in their optical spectra
  are classified as AGN-host galaxies,
  suggesting that AGNs may play a role in the evolution of these objects.

\end{enumerate}

Thanks to the wavelength coverage of Hectospec spectra,
  we identify several interesting features
  from the optical spectra of the 22 $\mu$m-selected galaxies:
  
\begin{enumerate}

\item The stacked spectra binned in IR luminosity
   show that the strength of the [O III] line relative to H$\beta$
   grows with increasing IR luminosity, which can
   suggest an increase in nuclear activity with IR luminosity.

\item We identify nine unusual galaxies with very strong
 [Ne III] $\lambda$3869, 3968 emission lines 
 (i.e. EW$_{\rm [Ne III] 3869}<-55$ \AA~and EW$_{\rm [Ne III] 3968}<-45$ \AA).
AGN and/or extremely young stars can produce this emission.
  
\end{enumerate}

We also study the effects of the presence of AGNs and
  of the nearest neighbor galaxy
  on the change in sSFRs of the 22 $\mu$m-selected galaxies:

\begin{enumerate}

\item The sSFR distribution of AGN-host galaxies
  is similar to pure SF galaxies,
  indicating the coexistence of SF and AGN.

\item The sSFRs of 22 $\mu$m-selected galaxies with 
  late-type neighbors appear to be larger than 
  those with early-type neighbors.
The fraction of galaxies with close neighbors
  increases with sSFR.
These results suggest an important role of the nearest neighbor galaxies
  in changing the SFA of galaxies.
  
\end{enumerate}

The combination of \wise data with 
  very deep, optical photometric data 
  provides a large sample of dusty galaxies;
we identify 48 DOG candidates 
  with large MIR to optical flux density ratios 
  (i.e. $S_{22\mu m}/S_{0.65 \mu m}\geq 982$).
None of these objects have SHELS spectroscopic data.
Ten of them have no optical counterparts even in the very deep, 
  DLS $R$-band source catalog.

Remarkably, \wise data probe the universe to $z\sim2$ (e.g., DOGs).
We identify these high-$z$ objects
  based on the very deep, DLS optical photometric data.
The SHELS dense spectroscopic survey data
  elucidate the spectroscopic properties 
  of the \wise sources at $z\lesssim0.8$.
To explore the nature of \wise sources further,
  we plan to investigate other fields
  including HectoMAP and DLS F1 field
  where there are extensive data sets
  including deep optical photometry, 
  dense redshift survey data, and uniform \wise photometry.

\acknowledgments

We thank the anonymous referee for his/her useful comments that
improved the original manuscript.
We thank Scott Kenyon for carefully reading the manuscript and Jong Chul Lee for useful discussion.
HSH acknowledges the Smithsonian Institution for the support of his post-doctoral fellowship.
The Smithsonian Institution also supports the research of MJG, MJK and DGF.
IPD is supported by NSF-AST grant AST-0708433.
Observations reported here were obtained at the MMT Observatory, a joint
  facility of the Smithsonian Institution and the University of Arizona.
This research has made use of the NASA/ IPAC Infrared Science Archive, 
which is operated by the Jet Propulsion Laboratory, California Institute of Technology, 
under contract with the National Aeronautics and Space Administration.
This publication makes use of data products from the Wide-field Infrared Survey Explorer, 
which is a joint project of the University of California, Los Angeles, 
and the Jet Propulsion Laboratory/California Institute of Technology, 
funded by the National Aeronautics and Space Administration.

\bibliographystyle{apj} 
\bibliography{ref_hshwang} 


\end{document}

%% file: table1.tex
\begin{deluxetable*}{crr}
\tablewidth{0pc} 
\tablecaption{Number of {\it WISE} 22 $\mu$m Sources in the SHELS Field
\label{tab-stat}}
\tablehead{
22 $\mu$m Sources & S/N$_{22 \mu m}\geq5$ & S/N$_{22 \mu m}\geq3$
}
\startdata
Total & 126 & 507\\
Sources with optical photometry ($R<25.2$) & 118 & 481\\
Sources with spectra ($R<21$) & 107 & 337\\
$D_n4000\geq1.5$ &   6 &  47\\
$D_n4000<1.5$ &  99 & 280\\
AGNs &  58 & 126\\
DOGs &   8 &  48\\
$[$Ne III$]$ Strong Galaxies &   5 &   9\\
\enddata
\end{deluxetable*}

%% file: table2.tex
\begin{deluxetable*}{cccccccc}
\tabletypesize{\footnotesize}
\tablewidth{0pc} 
\tablecaption{Optical Properties of {\it WISE} 22 $\mu$m-selected SHELS Galaxies\tablenotemark{a}
\label{tab-samp}}
\tablehead{
ID & SHELS ID & R.A.$_{2000}$ & Decl.$_{2000}$ & R (mag)     & z\tablenotemark{b} & $D_n4000$ & CLASS\tablenotemark{c}
}
\startdata
  1 & 138.7484675+30.9888536 &  9:14:59.63 & 30:59:19.87 &  19.503 & $0.40830\pm0.00011$ &  1.34 &  U \\
  2 & 138.7486030+29.7302448 &  9:14:59.66 & 29:43:48.88 &  12.543 & $0.02116\pm0.00000$ &  1.41 &  C \\
  3 & 138.7555069+29.8018154 &  9:15:01.32 & 29:48:06.54 &  18.792 & $0.18112\pm0.00006$ &  1.48 &  A \\
  4 & 138.7556940+29.0958824 &  9:15:01.37 & 29:05:45.18 &  20.335 & $0.33584\pm0.00006$ &  1.25 &  H \\
  5 & 138.7602425+29.6844158 &  9:15:02.46 & 29:41:03.90 &  18.122 & $0.13900\pm0.00010$ &  1.45 &  C \\
  6 & 138.7647624+29.2699887 &  9:15:03.54 & 29:16:11.96 &  13.739 & $0.02103\pm0.00006$ &  2.02 &  U \\
  7 & 138.7815804+29.6656783 &  9:15:07.58 & 29:39:56.44 &  17.523 & $0.18445\pm0.00010$ &  1.71 &  C \\
  8 & 138.7959984+30.1807242 &  9:15:11.04 & 30:10:50.61 &  18.945 & $0.09974\pm0.00009$ &  1.21 &  C \\
  9 & 138.7966319+29.2524034 &  9:15:11.19 & 29:15:08.65 &  15.764 & $0.02080\pm0.00000$ &  1.15 &  H \\
 10 & 138.8001139+30.0352458 &  9:15:12.03 & 30:02:06.88 &  16.335 & $0.12869\pm0.00008$ &  1.62 &  A \\
\enddata
\tablenotetext{1}{This table is available in its entirety in a machine-readable form in the online journal. A portion is shown here for guidance regarding its form and content.}
\tablenotetext{2}{The error of redshift is set to 0 when it is unavailable.}
\tablenotetext{3}{Galaxy classification (see \S \ref{agnsel}) : H (SF), C (Composite), A (AGN from optical spectra), M (AGN from MIR colors), B (Broad-line AGN), U (Undetermined)}
\end{deluxetable*}

%% file: table3.tex
\begin{deluxetable*}{ccrrrrrr}
\tabletypesize{\scriptsize}
\tablewidth{0pc} 
\tablecaption{{\it WISE} Properties of 22 $\mu$m-selected SHELS galaxies\tablenotemark{a}
\label{tab-wise}}
\tablehead{
ID & WISE ID & S$_{3.4}$ &  S$_{4.6}$ &  S$_{12}$ & S$_{22}$ & log($L_{IR}/L_\odot$) & log($M_{\rm star}/M_\odot)$ \\
   &         & (mJy)     &  (mJy)     &  (mJy)    &  (mJy)   &    &   
}
\startdata
  1 &  J091459.67+305919.9 & $    0.16\pm    0.01$ & $    0.10\pm    0.01$ & $    0.45\pm    0.28$ & $    2.89\pm    0.91$ &   11.75 &  11.40 \\
  2 &  J091459.66+294349.0 & $   27.36\pm    0.55$ & $   17.87\pm    0.33$ & $  104.99\pm    1.35$ & $  167.80\pm    3.25$ &   10.64 &  11.08 \\
  3 &  J091501.32+294806.4 & $    0.20\pm    0.01$ & $    0.29\pm    0.02$ & $    1.10\pm    0.13$ & $    4.59\pm    0.98$ &   11.08 &  10.64 \\
  4 &  J091501.41+290545.5 & $    0.03\pm    0.01$ & $    0.07\pm    0.01$ & $    0.46\pm    0.13$ & $    3.50\pm    0.93$ &   11.62 &  10.35 \\
  5 &  J091502.47+294103.8 & $    0.53\pm    0.02$ & $    0.42\pm    0.02$ & $    2.38\pm    0.15$ & $    6.99\pm    0.99$ &   10.99 &  10.89 \\
  6 &  J091503.55+291612.1 & $    5.62\pm    0.11$ & $    2.74\pm    0.06$ & $    1.67\pm    0.14$ & $    3.07\pm    0.95$ &    9.11 &  10.29 \\
  7 &  J091507.60+293956.1 & $    0.91\pm    0.02$ & $    0.70\pm    0.02$ & $    1.30\pm    0.14$ & $    3.96\pm    0.97$ &   11.04 &  11.42 \\
  8 &  J091511.05+301050.5 & $    0.06\pm    0.01$ & $    0.04\pm    0.01$ & $    0.31\pm    0.77$ & $    2.92\pm    0.86$ &   10.36 &   9.60 \\
  9 &  J091511.17+291508.1 & $    1.38\pm    0.05$ & $    0.94\pm    0.04$ & $    5.24\pm    0.25$ & $    6.84\pm    1.68$ &    9.42 &   9.58 \\
 10 &  J091512.03+300207.0 & $    1.17\pm    0.03$ & $    0.82\pm    0.03$ & $    2.56\pm    0.14$ & $    3.38\pm    0.88$ &   10.64 &  11.30 \\
\enddata
\tablenotetext{1}{This table is available in its entirety in a machine-readable form in the online journal. A portion is shown here for guidance regarding its form and content.}
\end{deluxetable*}

%% file: table4.tex
\begin{deluxetable*}{ccrrrrrl}
\tabletypesize{\scriptsize}
\tablewidth{0pc} 
\tablecaption{Dust-obscured galaxies in the SHELS field
\label{tab-dogs}}
\tablehead{
ID & WISE ID &  R\tablenotemark{a}  & S$_{3.4}$\tablenotemark{b} &  S$_{4.6}$ &  S$_{12}$ & S$_{22}$  & Comment \\
   &         &  (mag)         & (mJy)     &  (mJy)     &  (mJy)    &  (mJy)    & 
}
\startdata
  1 &  J091507.75+293811.9 &  23.064 & $    0.06\pm    0.01$ & $    0.17\pm    0.02$ & $    1.16\pm    0.14$ & $    3.05\pm    0.94 $ & Near edge of field                                                                         \\
  2 &  J091516.57+300001.0 &  99.999 & $    0.04\pm    0.01$ & $    0.03\pm    0.01$ & $    0.23\pm    0.00$ & $    2.51\pm    0.85 $ &                                                                                            \\
  3 &  J091518.50+290907.5 &  23.039 & $    0.06\pm    0.01$ & $    0.15\pm    0.01$ & $    0.48\pm    0.12$ & $    3.17\pm    0.88 $ & Two sources separated by 2\arcsec.5                                                        \\
  4 &  J091533.73+292025.2 &  22.790 & $    0.04\pm    0.01$ & $    0.04\pm    0.00$ & $    0.26\pm    0.00$ & $    2.91\pm    0.82 $ & Offset 0\arcsec.4 from {\it WISE} position                                                 \\
  5 &  J091534.16+301448.1 &  99.999 & $    0.03\pm    0.01$ & $    0.03\pm    0.00$ & $    0.23\pm    0.00$ & $    2.87\pm    0.82 $ &                                                                                            \\
  6 &  J091541.22+305624.7 &  23.015 & $    0.04\pm    0.01$ & $    0.06\pm    0.00$ & $    0.46\pm    0.14$ & $    3.64\pm    0.92 $ & Offset 1\arcsec.3 from {\it WISE} position                                                 \\
  7 &  J091546.49+300432.8 &  23.079 & $    0.01\pm    0.01$ & $    0.07\pm    0.01$ & $    0.47\pm    0.00$ & $    2.88\pm    0.95 $ & Offset 0\arcsec.5 from {\it WISE} position                                                 \\
  8 &  J091551.66+300627.8 &  24.386 & $    0.04\pm    0.01$ & $    0.05\pm    0.01$ & $    0.25\pm    0.00$ & $    3.06\pm    0.88 $ &                                                                                            \\
  9 &  J091609.21+294513.9 &  22.632 & $    0.05\pm    0.01$ & $    0.07\pm    0.01$ & $    0.25\pm    0.00$ & $    3.07\pm    0.87 $ &                                                                                            \\
 10 &  J091614.15+301944.8 &  22.421 & $    0.04\pm    0.01$ & $    0.03\pm    0.01$ & $    0.32\pm    0.14$ & $    4.00\pm    1.09 $ &                                                                                            \\
 11 &  J091630.98+294357.5 &  22.112 & $    0.09\pm    0.01$ & $    0.30\pm    0.02$ & $    2.62\pm    0.15$ & $    6.26\pm    1.14 $ & Point Source                                                                               \\
 12 &  J091636.07+300058.5 &  24.276 & $    0.04\pm    0.01$ & $    0.07\pm    0.01$ & $    0.24\pm    0.00$ & $    2.76\pm    0.90 $ &                                                                                            \\
 13 &  J091642.83+293613.1 &  22.049 & $    0.26\pm    0.01$ & $    0.51\pm    0.02$ & $    2.10\pm    0.14$ & $    5.74\pm    0.99 $ &                                                                                            \\
 14 &  J091647.60+304225.0 &  22.662 & $    0.08\pm    0.01$ & $    0.06\pm    0.01$ & $    0.27\pm    0.00$ & $    3.33\pm    0.96 $ &                                                                                            \\
 15 &  J091652.05+300553.0 &  22.997 & $    0.05\pm    0.01$ & $    0.10\pm    0.02$ & $    0.34\pm    0.13$ & $    2.99\pm    0.93 $ &                                                                                            \\
 16 &  J091654.90+294434.4 &  22.716 & $    0.06\pm    0.01$ & $    0.05\pm    0.01$ & $    0.30\pm    0.00$ & $    3.00\pm    0.91 $ &                                                                                            \\
 17 &  J091702.90+303921.0 &  99.999 & $    0.05\pm    0.01$ & $    0.08\pm    0.01$ & $    0.42\pm    0.00$ & $    2.98\pm    0.97 $ & Between three faint sources                                                                \\
 18 &  J091710.73+290309.8 &  22.465 & $    0.07\pm    0.01$ & $    0.08\pm    0.01$ & $    0.42\pm    0.00$ & $    3.24\pm    0.91 $ & Offset 0\arcsec.5 from {\it WISE} position                                                 \\
 19 &  J091716.01+300413.9 &  24.642 & $    0.04\pm    0.01$ & $    0.04\pm    0.00$ & $    0.50\pm    0.13$ & $    3.03\pm    0.93 $ & close pair with galaxy $2\arcsec$ away                                                     \\
 20 &  J091719.57+301253.9 &  99.999 & $    0.06\pm    0.01$ & $    0.05\pm    0.00$ & $    0.50\pm    0.00$ & $    3.25\pm    1.05 $ &                                                                                            \\
 21 &  J091731.51+303514.6 &  21.886 & $    0.09\pm    0.01$ & $    0.48\pm    0.02$ & $    5.21\pm    0.18$ & $   13.38\pm    1.07 $ &                                                                                            \\
 22 &  J091812.20+292045.6 &  99.999 & $    0.01\pm    0.00$ & $    0.05\pm    0.00$ & $    0.26\pm    0.00$ & $    4.86\pm    0.94 $ &                                                                                            \\
 23 &  J091830.86+303127.7 &  99.999 & $    0.43\pm    0.01$ & $    0.25\pm    0.02$ & $    0.37\pm    0.00$ & $    2.86\pm    0.91 $ &                                                                                            \\
 24 &  J091859.82+291218.8 &  23.516 & $    0.05\pm    0.01$ & $    0.10\pm    0.01$ & $    0.47\pm    0.00$ & $    3.67\pm    0.97 $ &                                                                                            \\
 25 &  J091904.71+290125.6 &  22.807 & $    0.07\pm    0.01$ & $    0.07\pm    0.02$ & $    0.43\pm    0.00$ & $    2.58\pm    0.86 $ & Companion galaxy $3\arcsec$ away                                                           \\
 26 &  J091911.74+293925.8 &  22.906 & $    0.05\pm    0.01$ & $    0.11\pm    0.02$ & $    0.49\pm    0.14$ & $    3.68\pm    0.96 $ & Three sources within $5\arcsec$ of {\it WISE} position                                     \\
 27 &  J091933.61+293838.7 &  23.143 & $    0.13\pm    0.01$ & $    0.31\pm    0.02$ & $    1.56\pm    0.14$ & $    3.63\pm    1.01 $ &                                                                                            \\
 28 &  J091938.12+300938.4 &  23.221 & $    0.09\pm    0.01$ & $    0.19\pm    0.02$ & $    1.39\pm    0.13$ & $    4.10\pm    1.00 $ & Multiple sources within 3\arcsec                                                           \\
 29 &  J091948.45+294242.9 &  23.226 & $    0.06\pm    0.01$ & $    0.09\pm    0.01$ & $    0.39\pm    0.13$ & $    2.94\pm    0.98 $ & Offset 1\arcsec.5 from {\it WISE} position                                                 \\
 30 &  J092018.07+302034.7 &  23.162 & $    0.04\pm    0.01$ & $    0.05\pm    0.01$ & $    0.34\pm    0.00$ & $    3.54\pm    0.96 $ & Offset 1\arcsec.2 from {\it WISE} position                                                 \\
 31 &  J092026.39+294817.4 &  23.508 & $    0.06\pm    0.01$ & $    0.04\pm    0.02$ & $    0.27\pm    0.00$ & $    3.25\pm    1.02 $ &                                                                                            \\
 32 &  J092029.31+294622.8 &  22.521 & $    0.06\pm    0.01$ & $    0.04\pm    0.02$ & $    0.37\pm    0.00$ & $    3.15\pm    0.99 $ &                                                                                            \\
 33 &  J092044.55+293824.4 &  22.749 & $    0.12\pm    0.01$ & $    0.09\pm    0.01$ & $    0.30\pm    0.13$ & $    3.34\pm    0.87 $ & Five sources within 5\arcsec                                                               \\
 34 &  J092048.18+290710.7 &  21.620 & $    0.16\pm    0.01$ & $    0.21\pm    0.02$ & $    2.11\pm    0.14$ & $   11.06\pm    1.07 $ & Offset $1\arcsec$ from {\it WISE} position                                                 \\
 35 &  J092100.52+293523.0 &  99.999 & $    0.08\pm    0.01$ & $    0.09\pm    0.01$ & $    0.24\pm    0.00$ & $    2.72\pm    0.89 $ &                                                                                            \\
 36 &  J092101.74+303140.5 &  99.999 & $    0.12\pm    0.01$ & $    0.09\pm    0.01$ & $    0.50\pm    0.00$ & $    3.74\pm    0.98 $ &                                                                                            \\
 37 &  J092105.59+301614.7 &  23.368 & $    0.11\pm    0.01$ & $    0.20\pm    0.02$ & $    1.57\pm    0.13$ & $    5.24\pm    0.92 $ &                                                                                            \\
 38 &  J092116.64+291946.2 &  24.740 & $    0.02\pm    0.01$ & $    0.08\pm    0.01$ & $    0.35\pm    0.14$ & $    2.97\pm    0.96 $ & Four sources within $4\arcsec$                                                             \\
 39 &  J092118.29+302910.2 &  23.477 & $    0.13\pm    0.01$ & $    0.17\pm    0.02$ & $    0.85\pm    0.13$ & $    3.35\pm    0.96 $ &                                                                                            \\
 40 &  J092129.47+291003.1 &  23.793 & $    0.04\pm    0.01$ & $    0.09\pm    0.01$ & $    1.79\pm    0.13$ & $    5.00\pm    0.93 $ &                                                                                            \\
 41 &  J092155.82+294154.5 &  22.798 & $    0.09\pm    0.01$ & $    0.09\pm    0.02$ & $    0.28\pm    0.00$ & $    3.04\pm    0.95 $ & Other source 2\arcsec.5 away                                                               \\
 42 &  J092209.52+292110.0 &  22.793 & $    0.04\pm    0.01$ & $    0.03\pm    0.01$ & $    0.34\pm    0.12$ & $    2.53\pm    0.83 $ &                                                                                            \\
 43 &  J092229.73+293254.0 &  23.145 & $    0.11\pm    0.01$ & $    0.26\pm    0.02$ & $    1.63\pm    0.14$ & $    3.71\pm    0.96 $ &                                                                                            \\
 44 &  J092311.14+302101.8 &  22.801 & $    0.13\pm    0.01$ & $    0.12\pm    0.02$ & $    0.96\pm    0.13$ & $    2.79\pm    0.91 $ & Three overlapping sources                                                                  \\
 45 &  J092322.42+301730.9 &  23.412 & $    0.15\pm    0.01$ & $    0.45\pm    0.02$ & $    1.91\pm    0.14$ & $    5.36\pm    1.01 $ &                                                                                            \\
 46 &  J092346.50+294250.0 &  23.420 & $    0.07\pm    0.01$ & $    0.21\pm    0.02$ & $    1.49\pm    0.13$ & $    3.57\pm    0.87 $ &                                                                                            \\
 47 &  J092352.59+303215.9 &  99.999 & $    0.05\pm    0.01$ & $    0.12\pm    0.01$ & $    0.42\pm    0.00$ & $    3.72\pm    0.91 $ &                                                                                            \\
 48 &  J092410.27+301716.8 &  99.999 & $    0.17\pm    0.01$ & $    0.14\pm    0.02$ & $    0.32\pm    0.00$ & $    3.21\pm    0.94 $ &                                                                                            \\
\enddata
\tablenotetext{1}{When there is no optical counterpart, the $R$-band magnitude is set to 99.999.}
\tablenotetext{2}{When the flux density error is equal to 0, the flux density indicates the upper limit.}
\end{deluxetable*}